\documentclass[journal,12pt,onecolumn,draftclsnofoot,]{IEEEtran}

\usepackage{amsthm,amsmath,amssymb}
\usepackage{mathrsfs}
\usepackage{amsfonts}
\usepackage{graphicx}
\usepackage{booktabs}
\usepackage{algorithm}
\usepackage{algorithmic}
\usepackage{graphics}
\usepackage{subfigure}
\usepackage{multirow}
\usepackage{bm}
\usepackage{cite}
\usepackage{diagbox}
\usepackage[table]{xcolor}
\usepackage{float}
\usepackage{makecell}

\ifCLASSINFOpdf
\else
\fi

\begin{document}

\title{Dual-Propagation-Feature Fusion Enhanced Neural CSI Compression for Massive MIMO}

\author{Shaoqing Zhang,
        Wei Xu,~\IEEEmembership{Senior Member,~IEEE}, Shi Jin,~\IEEEmembership{Senior Member,~IEEE}, Xiaohu You,~\IEEEmembership{Fellow,~IEEE}, Derrick Wing Kwan Ng,~\IEEEmembership{Fellow,~IEEE}, \\ and Li-Chun Wang,~\IEEEmembership{Fellow,~IEEE}
% <-this % stops a space
\thanks{S. Zhang, W. Xu, S. Jin, and X. You are with the National Mobile Communications Research Laboratory, Southeast University, Nanjing 210096, China, and also with Purple Mountain Laboratories, Nanjing 211111, China (e-mail: sq\_zhang@seu.edu.cn; wxu@seu.edu.cn; jinshi@seu.edu.cn; xhyu@seu.edu.cn). 

D. W. K. Ng is with the School of Electrical Engineering and Telecommunications, University of New South Wales, Sydney, NSW 2052, Australia (e-mail: w.k.ng@unsw.edu.au).

Li-Chun Wang is with the Department of Electrical Computer Engineering, National Yang Ming Chiao Tung University, Hsinchu 30010, Taiwan (e-mail: lichun@cc.nctu.edu.tw).}% <-this % stops a space
}
% \markboth{Journal of \LaTeX\ Class Files,~Vol.~14, No.~8, August~2015}%
% {Shell \MakeLowercase{\textit{et al.}}: Bare Demo of IEEEtran.cls for IEEE Journals}

\maketitle
\vspace{-1.6cm}
\begin{abstract}
Due to the ability of feature extraction, deep learning (DL)-based methods have been recently applied to channel state information (CSI) compression feedback in massive multiple-input multiple-output (MIMO) systems. Existing DL-based CSI compression methods are usually effective in extracting a certain type of features in the CSI. However, the CSI usually contains two types of propagation features, i.g., non-line-of-sight (NLOS) propagation-path feature and dominant propagation-path feature, especially in channel environments with rich scatterers. To fully extract the both propagation features and learn a dual-feature representation for CSI, this paper proposes a dual-feature-fusion neural network (NN), referred to as DuffinNet. The proposed DuffinNet adopts a parallel structure with a convolutional neural network (CNN) and an attention-empowered neural network (ANN) to respectively extract different features in the CSI, and then explores their interplay by a fusion NN. Built upon this proposed DuffinNet, a new encoder-decoder framework is developed, referred to as Duffin-CsiNet, for improving the end-to-end performance of CSI compression and reconstruction. To facilitate the application of Duffin-CsiNet in practice, this paper also presents a two-stage approach for codeword quantization of the CSI feedback. Besides, a transfer learning-based strategy is introduced to improve the generalization of Duffin-CsiNet, which enables the network to be applied to new propagation environments. Simulation results illustrate that the proposed Duffin-CsiNet noticeably outperforms the existing DL-based methods in terms of reconstruction performance, encoder complexity, and network convergence, validating the effectiveness of the proposed dual-feature fusion design.

\end{abstract}

\begin{IEEEkeywords}
Deep learning, massive multiple-input multiple-output (MIMO), channel state information (CSI), compression and reconstruction, dual-feature fusion.
\end{IEEEkeywords}

\IEEEpeerreviewmaketitle

\section{Introduction}
\IEEEPARstart
{I}{n} the last two decades, massive multiple-input multiple-output (MIMO) has received extensive attentions due to their great potential in improving the spectral efficiency (SE) of wireless communication networks \cite{6798744, 7386643}. As such, it serves as a core technology to enable the roll-out of the fifth-generation (5G) wireless communication networks \cite{wong2017key, 10024766}. In a practical massive MIMO system, the base station (BS) heavily relies on the availability of channel state information (CSI) for effective beamforming design to achieve high system SE and throughput \cite{7353214, wei2022distributed}. Indeed, it is crucial for the BS to acquire accurate CSI matrix in massive MIMO systems, especially in interference-limited multiuser communications. In time division duplexing (TDD) systems, the BS can directly estimate the uplink CSI through the pilot sequence transmitted by a user equipment (UE) and then use the estimate to predict the downlink CSI counterpart by exploiting channel reciprocity. However, currently deployed cellular networks are dominantly frequency division duplexing (FDD) systems. In particular, it becomes challenging to obtain the CSI in FDD as the channel reciprocity no longer holds. Therefore, it is necessary to feed the CSI back to the BS from a UE in FDD systems \cite{9279228}. However, for massive MIMO, the required feedback signaling overhead of CSI is scaled linearly with the increasing number of BSs antennas and active UEs, which is demonstrated by \cite{1715541} and \cite{7470522} from both the link level and network level, respectively, which calls for the design of effective CSI feedback, or reconsidering the role of the feedback channel \cite{5529760, wang2023full}.

Recently, due to the strong ability of feature extraction and cross-domain knowledge sharing, deep learning (DL) has been successfully applied in a wide range of fields in computer vision (CV) \cite{8766896}. Inspired by its success in the cross-field research, DL has attracted growing attention in wireless communication \cite{9295376, wei2023toward}. In particular, the advanced physical-layer technologies are triggered to redesign to exploit the potentials of DL, e.g., DL-based precoding designs \cite{1065,9246287,9729198, 9347820}, DL-enhanced channel estimation methods \cite{8400482,8795533,9288911}, DL-based MIMO detection approaches \cite{9075976, 9018199}, and DL-empowered security technologies \cite{9517121}. To unleash the potentials of DL, it has been introduced to the design of efficient CSI compression and reconstruction in systems requiring massive MIMO CSI feedback. Unlike conventional limited feedback methods \cite{7166317}, the studies in \cite{8322184, 9178295, 9149229} have revealed that DL-based CSI feedback methods have great potential in exploring the inherent structures of the CSI and offering superior performance thanks to the sparsity of massive MIMO channels. Specifically, in \cite{8322184}, inspired by the success of residual network (ResNet) \cite{He_2016_CVPR} in CV, an autoencoder network by integrating a residual convolutional neural network (CNN) and a fully-connected neural network (FNN), namely CsiNet, was first proposed for the CSI compression and reconstruction. It showed that the CsiNet outperformed some conventional methods, e.g., \cite{2004,7457256}, in terms of both the CSI reconstruction accuracy and the algorithm running time. On the basis of CsiNet, a series of effective techniques on massive MIMO CSI feedback are investigated. They mainly focused on the design of various NNs for improving the CSI reconstruction accuracy and satisfying practical needs.

On the one hand, considering the inherent image features of CSI matrices, an improved CNN-based CSI feedback scheme was proposed in \cite{9178295} for CSI compression by capturing the long-range dependencies in the CSI images. In addition, in \cite{9149229} and \cite{9495802}, two effective feedback CNNs and an advanced training scheme were proposed to improve the efficiency of CSI reconstruction by further considering multiple resolutions of CSI feature extraction. Also, in \cite{9373670}, a CNN with binary NN was deigned to improve the network performance and the speed of training convergence. Besides, in \cite{9552908}, an asymmetric convolution-based autoencoder framework was proposed for CSI feedback by utilizing asymmetric convolution blocks. The excellent performance of these existing CNN-based methods \cite{8543184, 9445070, lu2021binarized, 9419066, yzq} comes from the fact that CNNs can efficiently extract the non-line-of-sight (NLOS) propagation-path features in CSI images. However, in addition to these NLOS propagation-path features, there are small-region dominant propagation-path features in CSI images, which are not ignored in CSI compression and reconstruction with high accuracy. To focus a neural network (NN) on these dominant propagation-path features, an attention-empowered neural network (ANN) is recently proposed in \cite{9497358}. Compared to CNNs \cite{lu2021binarized, 9419066, yzq}, the proposed ANN effectively extracts the dominant propagation-path features in CSI images but it weakens the extraction of the NLOS propagation-path features, which results in the performance loss on CSI reconstruction.

On the other hand, many NN designs are proposed by considering practical needs and constraints, e.g., multi-rate NN \cite{8972904}, denoising NN \cite{9171358} and lightweight NN \cite{9439959}. In particular, to address the challenges of transmitting continuous codeword values, some DL-based feedback quantization designs have been proposed to apply the feedback quantization in network training to improve the performance of NNs with low-resolution feedbacks in practice \cite{9296555,9090892,8845636,9799802}. For instance, \cite{9296555} and \cite{9090892} proposed a noise injection method and a quantization approximation approach to apply the feedback quantization in the network training, respectively. Also, both the quantization gradient forgery method in \cite{8845636} and the quantization mapping method in \cite{9799802} can realize feedback quantization in network training.
% , which successfully improves the quantization performance. 
However, the quantization performance of these methods is limited by the adopted CSI feedback NN architecture. Besides, when the channel environment is varying, improving the generalization ability of the network is also a challenging problem.
% considering some practical needs and constraints, a multi-rate NN was proposed in \cite{8972904} to compress the CSI for not only improving the CSI reconstruction accuracy, but also bridging the gap between DL-based methods and their deployment in practice. Besides, in \cite{9171358}, a deep CNN architecture was proposed to conduct the CSI feedback by the inevitable errors in suppressing the impact of noisy CSI caused by practical channel estimation methods. To facilitate their implementation in wireless systems, a lightweight deep CNN for CSI compression and reconstruction was proposed in \cite{9439959} by exploiting the correlation characteristics of complex-valued channel responses in the angular-delay domain. 
% In addition, an integrated NN architecture including the designs of pilots, channel estimation, CSI feedback, and precoding calculation was proposed in \cite{9347820} to improve the end-to-end performance of the system. 
% In addition, some DL-based quantization designs have been proposed to involve the quantization into NN training to improve the low-resolution performance \cite{8845636,9296555,9090892,9799802,8461983,9057584}. 

Motivated by these facts, we propose a novel DL-based NN architecture with dual-feature fusion for CSI compression and consider the practical requirements in terms of codeword quantization and generalization enhancement. The major contributions of our work are summarized as follows:

\begin{sloppypar}
\begin{itemize}

\item We propose a novel dual-feature fusion NN framework for enhanced CSI processing, referred to as DuffinNet. The DuffinNet designs a parallel-serial hybrid architecture to fully extract physical features embedded in CSI images, achieving a better CSI feature representation compared to existing DL-based CSI compression NNs. Built upon DuffinNet, we design a new encoder-decoder NN framework, referred to as Duffin-CsiNet, for the massive MIMO CSI compression and reconstruction with high accuracy. It is the first work on dual-feature extraction and fusion in CSI compression and provides an ingenious and concise approach to leverage the complementary capabilities of CNN and ANN.

\item An adaptively weighted ANN based on autoencoder design is devised for dominant propagation-path feature extraction of CSI image in encoder. Correspondingly, a convolution-based ANN is adopted in decoder to reconstruct CSI. In the fusion stage, considering the differences between two CSI feature maps, we propose an NN-based fusion method for the fusion of the both feature maps extracted by NNs. On the other hand, a two-stage approach with the feedback quantization is applied for transmitting continuous codeword values and a transfer learning-based method is introduced to improve the generalization of the proposed Duffin-CsiNet.

\item We visualize the extracted feature map of the CSI image to verify the effectiveness of the proposed Duffin-CsiNet. In addition, two channel environments are used to discuss various performances of the network, including reconstruction performance, quantization performance, generalization performance, and network complexity. Experimental results show that the proposed Duffin-CsiNet outperforms existing DL-based CSI compression methods under five compression ratios. In particular, the reconstruction performance of the proposed Duffin-CsiNet achieves a normalized mean squared error (NMSE) of -35.19 dB under the compression ratio of 1/4, which has a 5.5 dB gain compared to existing state-of-the-art (SOTA) methods.

\end{itemize} 
\end{sloppypar}

The remainder of this paper is organized as follows. The system model and the problem formulation are introduced in Section II. Inherent features of CSI images and the design of DuffinNet are presented in Section III. Architecture, training and deployment of Duffin-CsiNet for CSI compression and reconstruction are elaborated in Section IV. Simulation results are provided in Section V. Conclusions are drawn in Section VI.

\emph{Notations}: Scalar variables, vectors, and matrices are denoted by lower case, boldface lower case, and boldface upper case letters, respectively. For a matrix $\mathbf{A}$, $\mathbf{A}^H$ and $\left\| \mathbf{A} \right\| $ denote its conjugate transpose and Frobenius norm, respectively. $\mathbb{E} \left\{ \cdot \right\}$ denotes the statistical expectation. $\mathcal{R} \left( \cdot \right) $ and $\mathcal{I} \left( \cdot \right) $ denote the real and imaginary part of a complex-valued variable, respectively. $a\times \mathbf{A}$ denotes scalar $a$ multiplies with every element of matrix $\mathbf{A}$. $\mathbf{b}\otimes \mathbf{A}$ denotes each element of vector $\mathbf{b}$ multiplies with corresponding channel matrix of matrix $\mathbf{A}$. $\mathbb{C} ^{m\times n}$ and $\mathbb{R} ^{m\times n}$ denote $m\times n$-dimensional complex matrices and real matrices, respectively. $\mathbb{C} ^{m\times n \times c}$ and $\mathbb{R} ^{m\times n \times c}$ denote $m\times n$-dimensional complex matrices and real matrices with $c$ image channels, respectively.

\section{System Model}
Consider an FDD massive MIMO communication system where the BS has $N_\mathrm{t}$ antennas and the UE equips $N_\mathrm{r}$ receiving antennas. Note that $N_\mathrm{t} \gg N_\mathrm{r}$ and $N_\mathrm{r}$ is set to 1 for simplicity, which is also popularly adopted by existing methods, e.g., \cite{9495802, 8543184, 9445070, 8972904, 9171358, 9439959, 9347820, lu2021binarized}. Orthogonal frequency division multiplexing (OFDM) with $N_\mathrm{c}$ subcarriers is employed. The received signal at the $n$-th subcarrier can be expressed as
\begin{equation}
    y_n=\mathbf{h}_{n}^{H}\mathbf{v}_nx_n+z_n, \ \ \forall n\in \left\{ 1,...,N_{\text{c}} \right\},
\end{equation}
where $x_n\in \mathbb{C}$ is the transmitted symbol, $z_n\in \mathbb{C}$ is the additive noise at the $n$-th subcarrier, and $\mathbf{h}_n\in \mathbb{C} ^{N_\mathrm{t}\times 1}$ and $\mathbf{v}_n\in \mathbb{C} ^{N_\mathrm{t}\times 1}$ denote the channel vector and the precoding vector corresponding to the $n$-th subcarrier, respectively. 

\begin{figure}[t]
    \centering
    \includegraphics[scale=0.55]{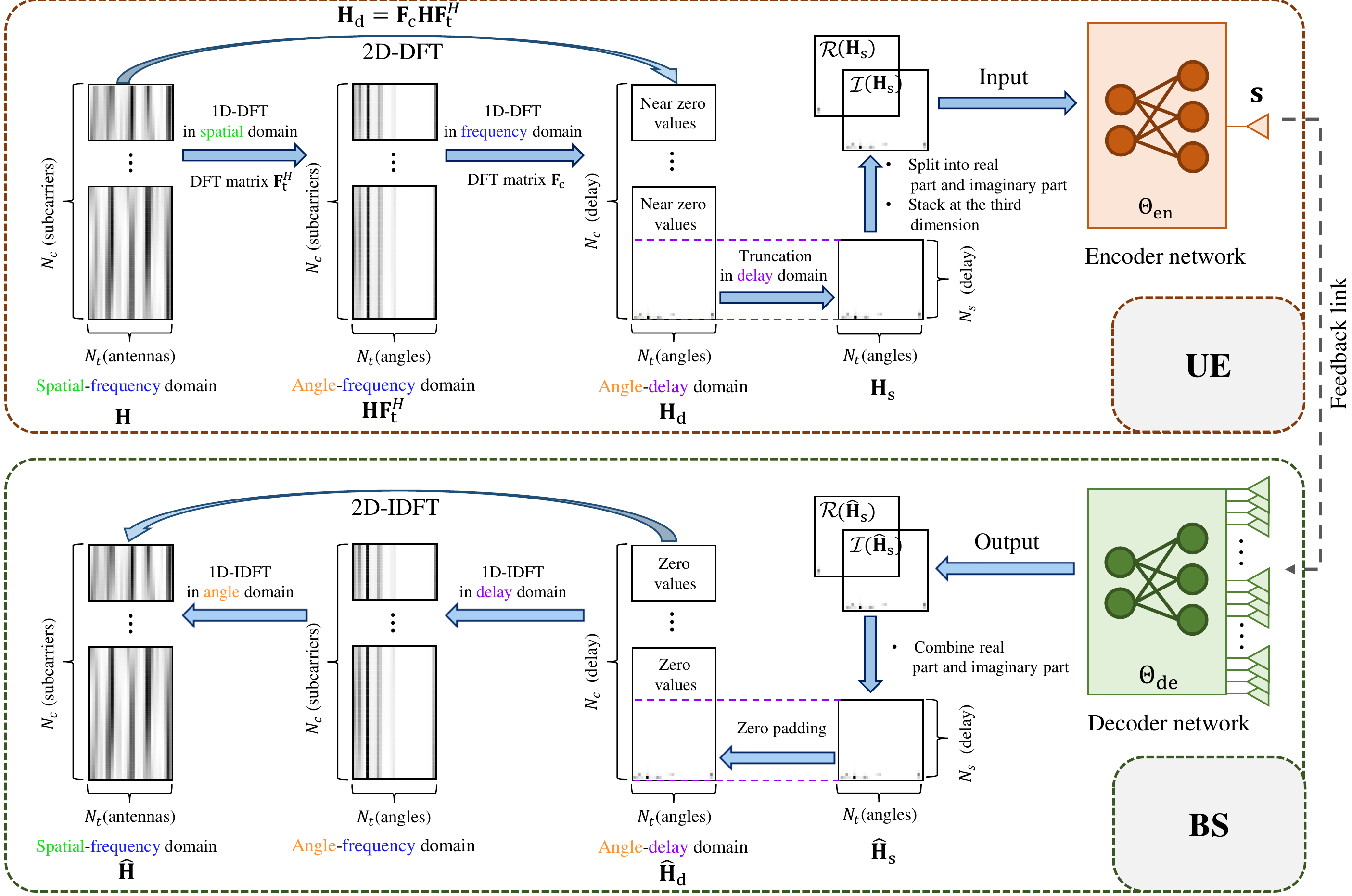}
    \caption{Procedure diagram of the CSI compression, feedback, and reconstruction.}
\end{figure} 

% \begin{figure}[t]
%     % \centering
%     \includegraphics[scale=0.5]{图WHC说明.pdf}
%     \caption{(a). RGB image (b). CSI image}
% \end{figure}

In the spatial-frequency domain, the complete CSI matrix at all the $N_\mathrm{c}$ subcarriers is denoted by $\mathbf{H}=\left[ \mathbf{h}_1,\cdots ,\mathbf{h}_{N_\mathrm{c}} \right] ^H\in \mathbb{C} ^{N_\mathrm{c}\times N_\mathrm{t}}$. Without compression, the number of total complex-valued feedback parameters, i.e., the size of $\mathbf{H}$, is $N_\mathrm{c}N_\mathrm{t}$, which is usually extremely large in massive MIMO systems. In order to reduce the feedback overhead, we have to first compress the CSI before feeding it back through a limited-capacity reverse link. To exploit the potential sparsity of $\mathbf{H}$ in the angular-delay domain in massive MIMO systems \cite{1033686}, we transform $\mathbf{H}$ from the spatial-frequency domain to the angular-delay domain before further processing. Using a two-dimensional (2D) discrete Fourier transform (DFT), the equivalent channel matrix in the angular-delay domain is expressed as 
\begin{equation}
    \mathbf{H}_\mathrm{d}=\mathbf{F}_\mathrm{c}\mathbf{HF}_{\mathrm{t}}^{H},
\end{equation}
where $\mathbf{F}_{\text{t}}^{H}\in \mathbb{C}^{N_{\text{t}}\times N_{\text{t}}}$ is a DFT matrix applied to the spatial domain for transforming the spatial-frequency-domain CSI matrix $\mathbf{H}$ to the angle-frequency-domain CSI matrix $\mathbf{H}\mathbf{F}_{\text{t}}^{H}$ and $\mathbf{F}_{\text{c}}\in \mathbb{C}^{N_{\text{c}}\times N_{\text{c}}}$ is a DFT matrix applied to the frequency domain for transforming the angle-frequency-domain CSI matrix to the angle-delay-domain CSI matrix $\mathbf{H}_\mathrm{d}$. The process of the 2D-DFT of $\mathbf{H}$ is shown at the upper half of Fig. 1. Consider the fact that practical multipaths arrive at limited delay intervals \cite{8845636}, $\mathbf{H}_\mathrm{d}$ only contains nonzero values in a short delay duration. Without loss of generality and following the same approaches as in \cite{9171358}, \cite{9439959}, we focus on the first $N_{\mathrm{s}}$ rows of $\mathbf{H}_\mathrm{d}$ in the angular-delay domain, denoted by $\mathbf{H}_\mathrm{s} \in \mathbb{C} ^{N_{\mathrm{s}}\times N_\mathrm{t}}$. In this manner, the number of parameters of the channel matrix decreases from $N_\mathrm{c}N_\mathrm{t}$ to $N_\mathrm{s}N_\mathrm{t}$.

To reduce the feedback signaling overhead and ensure accurate CSI reconstruction at the BS, DL-based methods have been applied for the CSI compression and reconstruction, e.g., \cite{9495802, 8543184, 9445070, 8972904, 9171358, 9439959, 9347820}. In particular, DL-based methods usually regard the CSI matrix as a multi-channel image mimicking a traditional visual image such that a deep neural network (DNN) can be applied for compression and feedback. Fig. 1 shows a typical procedure of the CSI compression feedback and reconstruction using DL. Specifically at the UE, the imaginary part and real part of $\mathbf{H}_\mathrm{s}$ form a 2-channel image, $\left[ \mathcal{R} \left( \mathbf{H}_\mathrm{s} \right) ;\mathcal{I} \left( \mathbf{H}_\mathrm{s} \right) \right]$, which is the input of the encoder network in this figure. Through the encoder network, the channel matrix is compressed into a short feature codeword vector, denoted by $\mathbf{s}$. Note that the length of $\mathbf{s}$ is determined according to the desired compression ratio $\rho \in (0, 1)$ of the system. Without loss of generality, the procedure of this DL-based CSI compression at the UE in Fig. 1 is represented by
\begin{equation}
    \mathbf{s}=f_{\mathrm{en}}\left( \left[ \mathcal{R} \left( \mathbf{H}_\mathrm{s} \right) ;\mathcal{I} \left( \mathbf{H}_\mathrm{s} \right) \right] ,\Theta _{\mathrm{en}} \right),
\end{equation}
where $f_{\mathrm{en}}\left( \cdot \right)$ denotes the encoder network and $\Theta _\mathrm{en}$ contains the training parameters of the encoder network. Afterwards, $\mathbf{s}$ is fed back to the BS such that the latter can use a decoder network to recover the truncated angle-delay-domain CSI matrix $\mathbf{H}_\mathrm{s}$ from the received codeword $\mathbf{s}$, which is represented by
\begin{equation}
    \hat{\mathbf{H}}_\mathrm{s}=f_{\mathrm{de}}\left( \mathbf{s},\Theta _{\mathrm{de}} \right) ,
\end{equation}
where $f_{\mathrm{de}}\left( \cdot \right)$ denotes the decoder network and $\Theta _\mathrm{de}$ contains the training parameters of the decoder network. Finally, conducting the zero padding on the $\hat{\mathbf{H}}_\mathrm{s}$ for obtaining the complete angle-delay-domain CSI matrix $\hat{\mathbf{H}}_\mathrm{d}$ and applying a two-dimensional inverse DFT (2D-IDFT)  on the $\hat{\mathbf{H}}_\mathrm{d}$ for reconstructing the spatial-frequency-domain $\mathbf{H}$, as shown at the bottom half of Fig. 1.

Our goal is to design the encoder network and decoder network, and train the parameters $\Theta _\mathrm{en}$ and $\Theta _\mathrm{de}$ such that the differences between $\mathbf{H}_\mathrm{s}$ and the recovered $\hat{\mathbf{H}}_\mathrm{s}$ are minimized. 

% Note that our work only focuses on the compression and feedback scheme, the downlink channel estimation is assumed to be ideal. Besides, we adopt COST2100 channle model \cite{6393523} to simulate the channel matrix $\mathbf{H}$ for the FDD massive MIMO system.

\section{CSI Features and Design of DuffinNet}
In this section, we introduce the the inherent physical features in CSI images and their corresponding feature extraction NNs popularly used. In particular, we visualize the CSI compression feature map extracted by existing NNs and discuss the feature extraction ability of existing CSI compression feedback NN architecture. In order to fully extract and fuse the physical features in CSI images, we propose the dual-feature fusion enhanced NN, which consists of three new functional networks.

\begin{figure*}[t]
    \centering
    \includegraphics[scale=0.45]{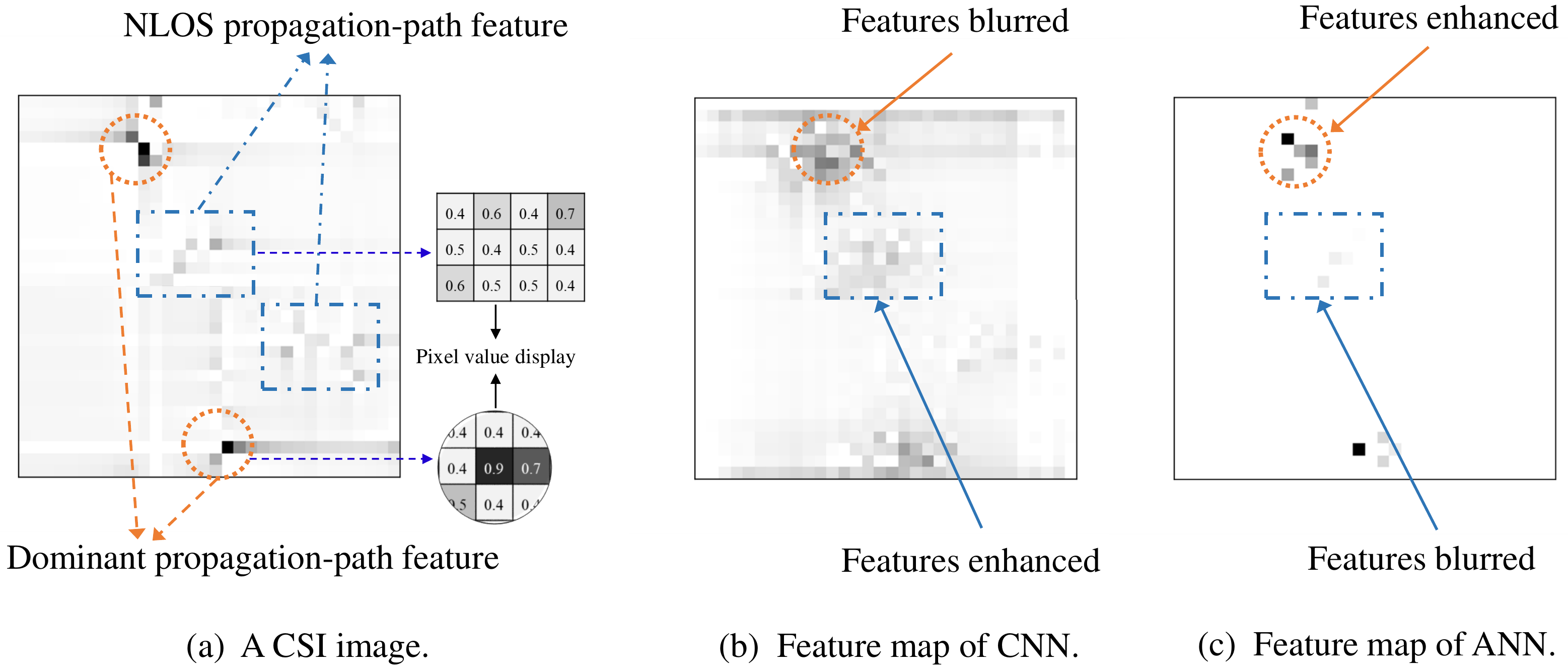}
    \caption{Inherent physical features in a CSI image.}
\end{figure*}

\subsection{Inherent Physical Features in CSI Images}
% In general, due to the randomness of the multipath fading and the user locations, the inherent features of a CSI image can be divided into two classes: gradual features and sharp features \cite{yzq}.

%  In this figure, each pixel is the signal strength of $i$-subcarrier received by $j$-UE \cite{8322184}. 
In practice, due to the randomness of multipath fading and UE locations \cite{lu2021binarized}, CSI images usually contain few sharp pixels with large values, which can be observed by an example diagram of the CSI image with pixel values displayed, as shown in Fig. 2(a). In particular, they have large absolute values and are significantly different from their neighboring pixels, which usually represents the dominant propagation path in practical systems. The spatial pattern contained in these sharp pixels is regarded as the dominant propagation-path features of the CSI images. For pixels other than sharp pixels, they usually carry small absolute values and tend to repeat themselves with slight and smooth changes in the neighboring regions, which denotes the propagation path with scatterers and reflectors in practical multipath environment. Similarly, this spatial pattern with the smooth nature is regarded as the NLOS propagation-path features of the CSI images. Fig. 2(a) is a typical CSI image from the COST2100 channel model \cite{6393523}. In particular, this image refers to the real part of the truncated angle-delay-domain CSI image $\mathbf{H}_\mathrm{s}$, i.e., the operation of 2D-DFT and the removing of zero values have been conducted. By observing the CSI image, it can be found that the NLOS propagation-path features are widespread compared with the dominant propagation-path features, as only a few strong paths exists in the multipath and the others are weak paths in practical communication environment.

% Obviously, the effective extraction of the gradual features plays an important role in improving the performance of CSI reconstruction thanks to their dominance in numbers, which has been verified in, e.g., \cite{9495802, 8543184, 9445070, 8972904, 9171358, 9439959, 9347820}. However, in practice, the sharp features are more important than the gradual features as they contain stronger received subcarrier signals, which is crucial for the quality of CSI-based techniques, e.g., precoding design at BS. Although the number of sharp features is small, solely focusing on the gradual features is performance-limiting in terms of CSI reconstruction accuracy and the performance of CSI-based techniques. Therefore, the processing of the sharp features is also urgently needed. 

For handling the NLOS propagation-path features in CSI image, typical CNNs are usually adopted, e.g., in \cite{9445070, 8972904, 9171358}, to extract them. As shown in Fig. 3(a), the CNN-based encoder is first used to extract the NLOS propagation-path feature map and then compress it into a codeword vector for feedback according to the compression ratio, and finally the corresponding CNN-based decoder is designed to reconstruct the CSI image from the received codeword. Fig. 2(b) shows the extracted feature map of the CSI image by CNNs. It is observed that the CNNs extract a feature map that enhances the NLOS propagation-path features for compression, as this similar regional features with smooth nature are easily captured by convolutional operations \cite{yzq}. However, the challenge in designing effective CNN-based CSI feedback is that CNNs usually blur the important information related to the dominant propagation path, i.e., the dominant propagation-path features. In particular, the dominant propagation path is the path between the transmitter and receiver that exhibits the strongest signal strength and is critical to the overall performance in wireless communication systems \cite{7368949}. It is used to estimate the channel parameters, e.g., signal-to-noise ratio, channel capacity, channel quality, etc., which is crucial for physical-layer technologies such as beamforming and signal detection. 

\begin{figure*}[t]
    \centering
    \includegraphics[scale=0.4]{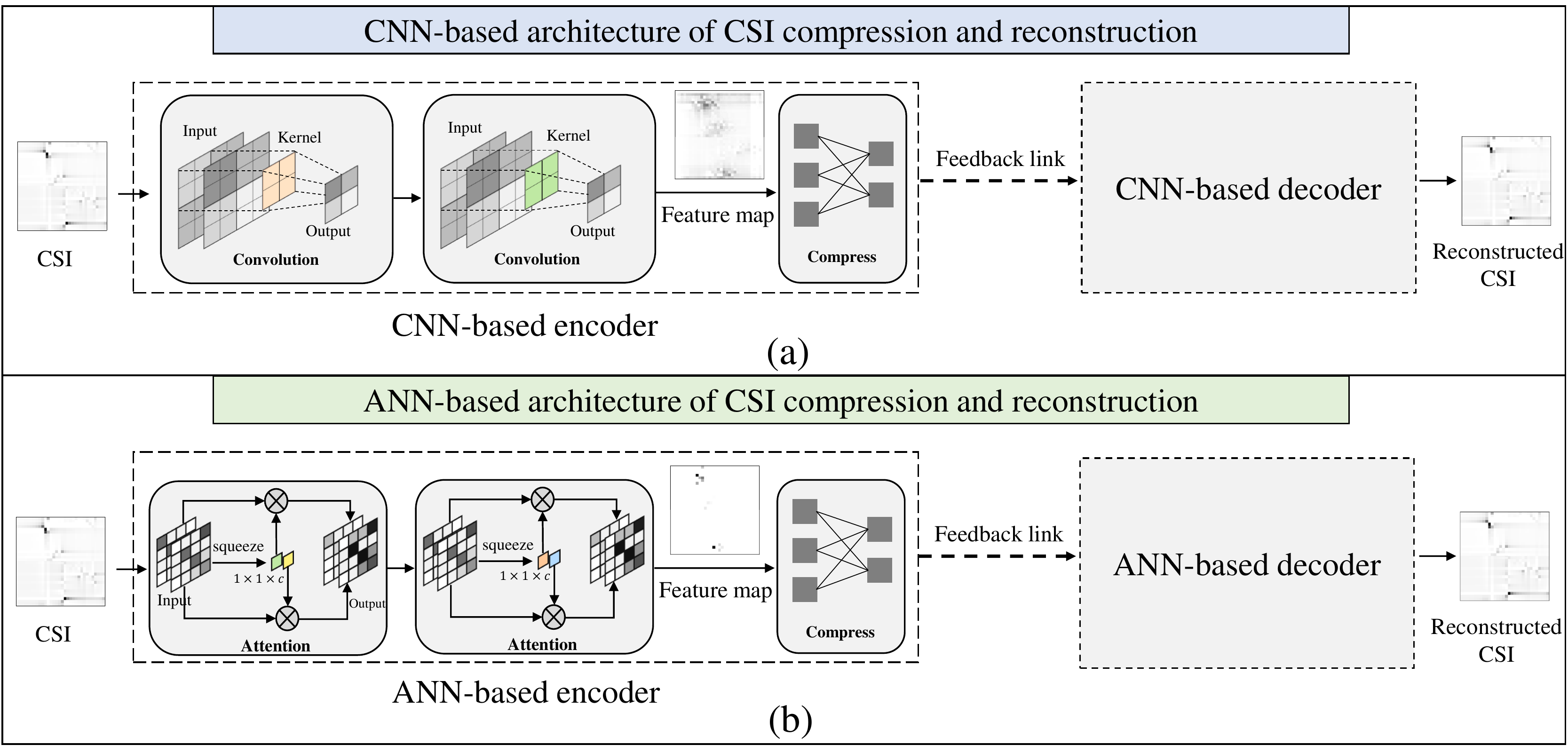}
    \caption{(a). Existing CNN-based CSI feedback architecture; (b). Existing ANN-based CSI feedback architecture.}
\end{figure*}

In order for the NNs to effectively focus on the dominant propagation-path features of a CSI image, attention mechanism have been proposed, e.g., in \cite{woo2018cbam, 9156697}, and some attention-empowered NNs (ANNs) are developed for CSI compression, e.g., in \cite{9497358}. A typical diagram of an ANN-based CSI compression and reconstruction is shown Fig. 3(b). Different from the convolution operation of CNNs, the ANNs first squeeze the input CSI image into a weight value vector with a dimension of $1 \times 1 \times c$, where $c$ denotes the number of the input CSI image channel, which is generally implemented by a pooling-based FNN. Specifically, the pooling operation, e.g., average pooling and max pooling, can help NNs to focus the receptive field on the large-valued regions where spatial pattern exhibits sharp nature. Then, through an FNN, an attention vector, i.e., the weight value vector, can be learned, which represents the importance of the dominant propagation-path features on all the CSI image channels. Finally, the element-wise product between the squeezed weight value vector and the input image is performed to obtain a feature map, which selectively highlights the dominant propagation-path features on different image channels. In particular, the dominant propagation-path features with large weights on the dimension of CSI image channels play a more significant role for improving the performance of CSI reconstruction \cite{9156697}. Fig. 2(c) shows a feature map extracted by the ANNs. We observe that ANNs extract a feature map that greatly enhances the dominant propagation-path features of the original CSI image, but dilute the extraction of the NLOS propagation-path features in the CSI image.

In fact, for NLOS propagation paths, they also play an essential role in the CSI feedback, which provides information about the obstacles affecting the signal transmission in wireless environment. By capturing the NLOS propagation-path features in the CSI feedback, the transmitter can perform efficient signal processing to minimize the interference and fading caused by the NLOS paths \cite{7368949}. As a result, both the dominant and NLOS propagation-path features need to be considered in the design of CSI feedback networks. Motivated by these observations and analysis, i.e., 
% Based on the above observations and analysis, it can be seen that although the existing DL-based CSI compression networks have excellent reconstruction performance, they are still limited by their feature extraction ability. Therefore, motivated by
the need of effective extraction of dual features in the CSI compression, we propose a dual-feature fusion enhanced network architecture, i.e., DuffinNet, for the processing of the CSI image. In particular, for the fusion of different CSI feature maps, instead of simply performing element-wise addition or dot product on them, we design a concise NN for efficient fusion.

\begin{figure*}[t]
    \centering
    \includegraphics[scale=0.48]{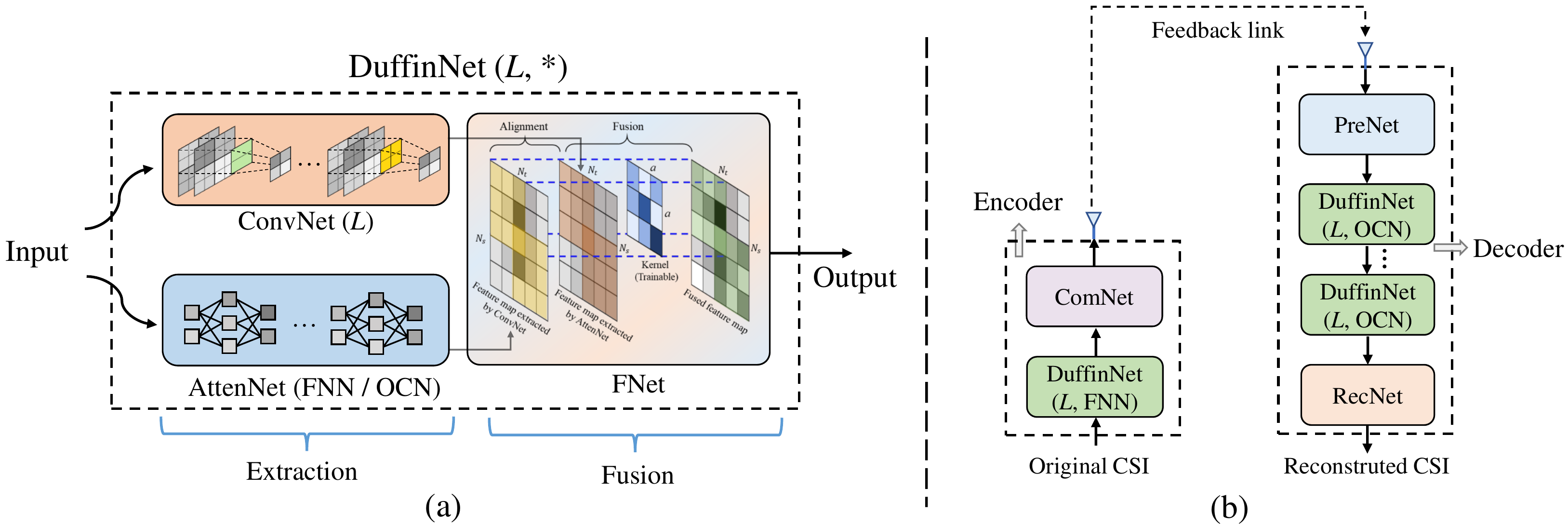}
    \caption{(a) Architecture of the proposed DuffinNet, where $L$ denotes the layer number of ConvNet and ``$*$'' represents FNN or OCN; (b) Framework of the proposed Duffin-CsiNet.}
\end{figure*}

% \begin{figure}[t]
%     \centering
%     \includegraphics[scale=0.53]{CNN与ANN.pdf}
%     \caption{The diagrams of the typical CNN and ANN.}
% \end{figure}

% \begin{figure*}[t]
%     \centering
%     \includegraphics[scale=0.53]{CSI压缩恢复整体图.pdf}
%     \caption{Architecture of the proposed Duffin-CsiNet.}
% \end{figure*}

\subsection{DuffinNet Architecture}
A concise architecture of the proposed DuffinNet is shown in Fig. 4(a). In particular, it is a hybrid parallel-serial structure and consists of three networks: convolutional network (ConvNet), attention-empowered network (AttenNet), and fusion network (FNet). The ConvNet is responsible for the extraction of the NLOS propagation-path features embedded in the CSI. It is composed of $L$ convolutional layers with different convolutional kernels. Different from the ConvNet, the AttenNet focuses on extracting and manipulating the dominant propagation-path features of the CSI images. Specifically, it consists of either an FNN or an one-dimensional (1D) convolutional network (OCN) with the attention mechanism. Note that the number of convolutional layers, $L$, and the structure selection of AttenNet are regarded as configurable parameters of the DuffinNet. This parallel network structure successfully unleashes the potential of different types of NNs in terms of extracting different inherent features of an image. To fully leverage the complementary strengths between the dominant propagation-path feature map derived from the ConvNet and the NLOS propagation-path feature map obtained from the AttenNet, we design a concise fusion network, i.e., the FNet in Fig. 4(a), to integrate both feature maps for learning a more representative feature map of CSI image. The FNet first concatenates the both feature maps in the image-channel dimension, i.e., the third dimension, which guarantees that each pixel of the two feature maps is aligned. Then, a trainable convolutional kernel with a size of $a\times a$ is utilized to perform a weighted fusion on the pixels in the $a\times a$ region of the both feature maps. Each weight, representing the relative importance of each pixel in the kernel, is dynamically adjusted by the neural network during the stage of training. By sliding the kernel over the concatenated feature map, FNet produces a more representative feature map that effectively fuses the both features of the CSI image for compression. This NN-based fusion approach achieves an efficient fusion by introducing trainable weight variables to explore the importance level of each feature map, providing a concise and effective scheme for the fusion involving CSI in wireless communication.

% Rather than using a direct addition, this fusion method introduces a learnable parameter, i.e. the $1 \times 1$ convolution kernel, which can fully explore the correlation between different feature maps and further improve the learning ability of the network.

Built upon the proposed DuffinNet, we are ready to develop an encoder-decoder network, i.e., Duffin-CsiNet, for the MIMO CSI compression and reconstruction. The architecture of the proposed Duffin-CsiNet is shown in Fig. 4(b). The overall component of DuffinNet is applied in both the encoder and decoder of the Duffin-CsiNet. In particular, the DuffinNet in the encoder of Duffin-CsiNet is used to extract the inherent physical features of the CSI images for compression where FNN-based AttenNet is adopted, while the DuffinNet in the decoder of Duffin-CsiNet is responsible for the feature procession of the compressed codeword vector for CSI reconstruction where OCN-based AttenNet is used. The design details are described in the following sections. Note that similar to previous works, e.g., \cite{9495802, 8543184, 9445070, lu2021binarized}, the design of NNs is focused and the feedback error is not considered in this paper.

\section{Design of Duffin-CsiNet}
In this section, we elaborate the design of Duffin-CsiNet as sketched in Fig. 4(b). In particular, we first introduce the design of the encoder network, then present the structure of the decoder network, and finally illustrate the training and deployment of Duffin-CsiNet. 
% the encoder is used to compress the CSI image and consists of two component networks: DuffinNet-E and compression network (ComNet). Then the compressed codeword vector is fed back to the BS from the feedback link. At the BS, an DuffinNet-based decoder is used to restore the CSI image from the received codeword vector and it consists of three sub-networks: preprocessing network (PreNet), cascaded DuffinNet-D, reconstruction network (RecNet). Detailed elaboration of each module of the Duffin-CsiNet is as follows.

\subsection{Encoder of Duffin-CsiNet}
% The encoder in Duffin-CsiNet is used to compress the CSI image and consists of two component networks: DuffinNet-E and ComNet. The DuffinNet-E is used to extract the inherent features of the CSI image and consists of three sub-networks: ConvNet, adaptive fusion-based AttenNet, and AFNet. The ComNet is to compress the extracted feature maps into codeword vector according to the specific compression ratio. The detailed structure of the encoder is shown in Fig. 4(b). As a crucial component of the encoder, each sub network of DuffinNet-E is introduced one by one as follows

The detailed structure of the encoder network is shown in Fig. 5(a). It consists of two parts: the DuffinNet and the compression network (ComNet). When we input an original CSI image into the encoder network, the DuffinNet is the first processor which fully extracts the inherent features embedded in the CSI image. The DuffinNet in the encoder consists of three components, i.e., an $L$-layer ConvNet, an FNN-based AttenNet, and an FNet.

In particular, the ConvNet is used to extract the NLOS propagation-path feature maps of the CSI image and is composed of $L$ composite convolutional layers with different convolutional kernels. Each composite convolutional layer is a proper combination of a convolutional  layer, a batch normalization layer, and an activation layer \cite{9171358}. 
% convolution layer $l=1,\dots,L$ creats $k_l$ convolution kernels of size $a_{l}^{1}\times a_{l}^{2}$. 
In the $l$-th composite convolutional layer, let $\mathbf{O}_{l}^{\mathrm{c}}\in \mathbb{R} ^{h_l\times w_l\times k_l}$, $\mathbf{O}_{l}^{\mathrm{b}}\in \mathbb{R} ^{h_l\times w_l\times k_l}$, and $\mathbf{O}_{l}^{\mathrm{a}}\in \mathbb{R} ^{h_l\times w_l\times k_l}$ denote the corresponding outputs of the convolutional layer, the batch normalization layer, and the activation layer, respectively, where $l \in \{1,\dots,L\}$ and $h_l$, $w_l$, and $k_l$ are the height, the width, and the number of channel of the output image, respectively. Denote the computations involved in the $l$-th composite convolutional layer by
$\mathbf{ComConv}_l\left( \mathbf{O}_{l-1}^{\mathrm{a}} , a_{l}^{1}\times a_{l}^{2}\times k_l \right)$, which consists of the following three steps
\begin{align}
&\mathbf{O}_{l}^{\mathrm{c}}=\mathbf{Conv2d}\left( \mathbf{O}_{l-1}^{\mathrm{a}},\mathbf{\Theta }_{l}^{\mathrm{c}},\boldsymbol{\epsilon }_{l}^{\mathrm{c}},a_{l}^{1}\times a_{l}^{2}\times k_l \right),
\\
&\mathbf{O}_{l,k}^{\mathrm{b}}\left[ m,n \right] =\frac{\mathbf{O}_{l,k}^{\mathrm{c}}\left[ m,n \right] -\phi _{l,k}}{\sqrt{\varepsilon _{l,k}+\zeta}},
\\
&\mathbf{O}_{l}^{\mathrm{a}}=\sigma\left( \mathbf{O}_{l}^{\mathrm{b}} \right),
\end{align}
where the operator $\mathbf{Conv2d}$ denotes a 2D convolutional operation and $\mathbf{\Theta }_{l}^{\mathrm{c}}\in \mathbb{R} ^{a_l\times a_l\times k_l}$ and $\boldsymbol{\epsilon }_{l}^{\mathrm{c}}\in \mathbb{R} ^{k_l\times 1}$ are the weights and bias vector of the convolutional kernels in the $l$-th composite convolutional layer, respectively. $\mathbf{O}_{l,k}^{\mathrm{b}}\left[ m,n \right]$ and $\mathbf{O}_{l,k}^{\mathrm{c}}\left[ m,n \right]$, $m\in \{1,\dots,h_l\}$, $n\in \{1,\dots,w_l\}$, denote the $(m,n)$-th element of the $k$-th image channel, $k\in \{1,\dots,k_l\}$, of $\mathbf{O}_{l}^{\mathrm{b}}$ and $\mathbf{O}_{l}^{\mathrm{c}}$, respectively, $\phi _{l,k}$ and $\varepsilon _{l,k}$ are the batch mean and variance of the $k$-th image channel of $\mathbf{O}_{l}^{\mathrm{c}}$, respectively, and $\zeta$ is a small float added to the variance to avoid dividing by zero. $\sigma(\cdot)$ denotes a nonlinear activation function and here the Leaky Rectified Linear Unit (LeakyReLU) function is applied
\begin{equation}
    \mathrm{LeakyReLU}\left( x \right)=\begin{cases}
	x,&		x\geqslant 0,\\
	\alpha x,&		x<0,\\
\end{cases}
\end{equation}
where $\alpha$ is an adjustable super parameter. Note that $\mathbf{O}_{0}^{\mathrm{a}} \in \mathbb{R} ^{N_\mathrm{t}\times N_\mathrm{s}\times 2}$ is the input CSI image and $\mathbf{O}_{L}^{\mathrm{a}} \in \mathbb{R} ^{N_\mathrm{t}\times N_\mathrm{s}\times 2}$ is the extracted NLOS propagation-path feature maps. We use $\mathbf{G}$ to represent $\mathbf{O}_{L}^{\mathrm{a}}$ in the later description. Compared to the convolutional layer, the composite convolutional layer can reduce the possibility of over-fitting, accelerate the speed of convergence, and make the NN less sensitive to the initialization of weights \cite{8935405}.

\begin{figure*}[t]
    \centering
    \includegraphics[scale=0.45]{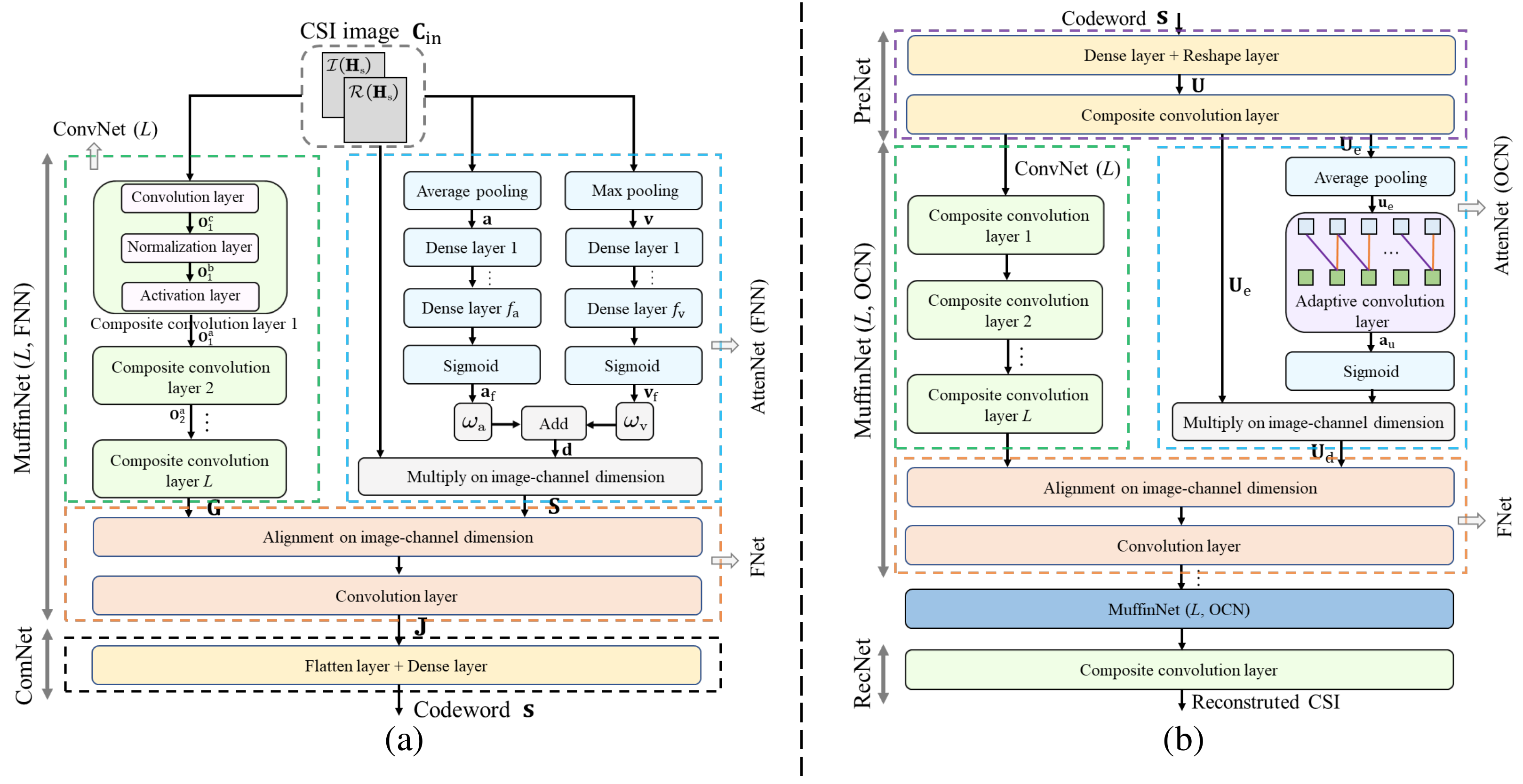}
    \caption{(a). The structure of the encoder in Duffin-CsiNet; (b). The structure of the decoder in Duffin-CsiNet.}
\end{figure*}

% \begin{figure*}[t]
%     \centering
%     \includegraphics[scale=0.48]{DuffinNet结构图与CSI压缩恢复整体图.pdf}
%     \caption{(a) Architecture of the proposed DuffinNet, where ``$L$'' denotes the layer number of ConvNet and ``$*$'' represents FNN or OCN; (b) Framework of the proposed Duffin-CsiNet.}
% \end{figure*}
% Besides, with the deepening of composite convolution layer, more significant sharp features can be extracted.

On the other hand, the FNN-based AttenNet is used to force the NN to focus on the dominant propagation-path features of the CSI image. As shown the blue dotted box in Fig. 5(a), it consists of three sub-networks: Average pooling-based FNN, max pooling-based FNN, and an adaptively weighted network. Specifically, the input CSI image is first processed by two pooling layers. The function of the pooling operation is to remove the redundant information, i.e., the small-valued pixels, and let the NN focus on the large-valued pixels, i.e., the dominant propagation path region of CSI image. In particular, inspired by the work in \cite{woo2018cbam}, we adopt two common pooling operations, i.e., the average pooling and the max pooling, to fully explore the dominant propagation-path features of the CSI image for effective compression.
Through the both pooling layers, the average value-based vector, denoted as $\mathbf{a} \in \mathbb{R} ^{1\times 1\times 2}$, and the max value-based vector, denoted as $\mathbf{v} \in \mathbb{R} ^{1\times 1\times 2}$, are obtained, respectively, i.e., 
\begin{align}
    &\mathbf{a}\left[ k \right] =\frac{\sum_{i=1,\ j=1}^{i=N_{\text{s}},\ j=N_{\text{t}}}{\text{C}_{\text{in,}k}\left[ i,j \right]}}{N_{\text{s}}N_{\text{t}}},
    \\
    &\mathbf{v}\left[ k \right] =\mathrm{max}\left( \mathbf{C}_{\text{in,}k} \right), 
\end{align}
where $\mathbf{C}_{\mathrm{in}}\in \mathbb{R} ^{N_{\mathrm{s}}\times N_\mathrm{t}\times 2}$ is the input of encoder, i.e., the CSI image, ${\mathbf{C}_{\text{in,}k}\left[ i,j \right]}$ denotes the $(i,j)$-th element of the $k$-th image channel of $\mathbf{C}_{\text{in}}$, and the max operator here is used to take the maximum value in the $k$-th image channel of $\mathbf{C}_{\text{in}}$. 
Then, two attention vectors, denoted as $\mathbf{a}_{\mathrm{f}} \in \mathbb{R} ^{1\times 1\times 2}$ and $\mathbf{v}_{\mathrm{f}} \in \mathbb{R} ^{1\times 1\times 2}$, are further learned by two FNNs, respectively. The calculations involved in the two FNNs are expressed as
\begin{align}
    &\mathbf{a}_{\mathrm{f}}=\mathbf{\Gamma }_{f_\mathrm{a}}\left( \mathbf{\Gamma }_{f_\mathrm{a}-1}\cdots \left( \mathbf{\Gamma }_1\mathbf{a}+\mathbf{\tau }_1 \right) \cdots +\mathbf{\tau }_{f_\mathrm{a}-1} \right) +\mathbf{\tau }_{f_\mathrm{a}},
    \\
    &\mathbf{v}_{\mathrm{f}}=\mathbf{\Lambda }_{f_\mathrm{v}}\left( \mathbf{\Lambda }_{f_\mathrm{v}-1}\cdots \left( \mathbf{\Lambda }_1\mathbf{v}+\mathbf{\pi }_1 \right) \cdots +\mathbf{\pi }_{f_\mathrm{v}-1} \right) +\mathbf{\pi }_{f_\mathrm{v}},
\end{align}
respectively, where $f_\mathrm{a}$ and $f_\mathrm{v}$ are the number of layer of the both FNNs, respectively. $\mathbf{\Gamma }_i\in \mathbb{R} ^{p_i\times p_{i-1}}$ and $\mathbf{\tau }_i\in \mathbb{R} ^{p_i\times 1}$, $i\in \{1,2,\cdots ,f_\mathrm{a}\}$, account for the weight matrix and bias vector of $i$-th fully-connected layer with $p_i$ neurons after average pooling layer, respectively, and $\mathbf{\Lambda }_j\in \mathbb{R} ^{q_j\times q_{j-1}}$ and $\mathbf{\pi }_j\in \mathbb{R} ^{q_j\times 1}$, $j\in \{1,2,\cdots ,f_\mathrm{v}\}$, are the weight matrix and bias vector of $j$-th fully-connected layer with $q_j$ neurons after max pooling layer, respectively. Note that in order to speed up the network convergence and avoid potential over fitting, both batch normalization layer and activation function layer are applied after each fully-connected layer. In particular, the activation function in the last fully-connected layer adopts the Sigmoid function to limit the values of the attention vector to [0, 1]. The Sigmoid function is given by 
\begin{equation}
    \mathrm{Sigmoid}\left( x \right) =\frac{1}{1+e^{-x}},
\end{equation}
and the other layers usually use the Rectified Linear Unit (ReLU) function, i.e., 
\begin{equation}
    \mathrm{ReLU}\left( x \right) =\max \left( 0,x \right).
\end{equation}

% \begin{figure}[t]
%     \includegraphics[scale=0.43]{Encoder结构.pdf}
%     \caption{The structure of the encoder in Duffin-CsiNet.}
% \end{figure}

After obtaining both the attention vectors, an adaptively weighted addition strategy is proposed to effectively acquire the final attention vector. Specifically, we introduce two trainable weights, denoted as $\omega_{\mathrm{a}}$ and $\omega_{\mathrm{v}}$, and multiply them with the both attention vectors respectively to obtain the final attention vector, denoted by $\mathbf{d} \in \mathbb{R} ^ {1 \times 1 \times 2}$, which is expressed as
\begin{equation}
    \mathbf{d}=\omega _\mathrm{a}\times \mathbf{a}_{\mathrm{f}}+\omega _\mathrm{v}\times \mathbf{v}_{\mathrm{f}}.
\end{equation}

Note that at the beginning of network training, we need to initialize $\omega _\mathrm{a}$ and $\omega _\mathrm{v}$, and then they are involved in network training as network trainable parameters, i.e., $\omega _\mathrm{a}$ and $\omega _\mathrm{v}$ are adjusted adaptively by the proposed network. This adaptive approach fully releases the learning ability of the NN by introducing two trainable variables, which promotes the NN to obtain an attention vector that effectively represents the importance of dominant propagation-path features on all image channels, especially in the case of no prior experience with the input image.
% the proportion of important elements of attention vector, especially in the case of no prior experience with the input image. 
% In addition, different initializations of $\omega _\mathrm{a}$ and $\omega _\mathrm{v}$ would have different impact on the performance of the network, which will be discussed in the simulation analysis in Section V.

Finally, the dominant propagation-path feature maps, denoted as $\mathbf{S} \in \mathbb{R} ^{N_\mathrm{s}\times N_\mathrm{t}\times 2}$, is obtained by multiplying $\mathbf{d}$ with the original CSI image, i.e., $\mathbf{C}_{\mathrm{in}}$, on the image-channel dimension, which is expressed as 
\begin{equation}
    \mathbf{S}=\mathbf{d}\otimes \mathbf{C}_{\mathrm{in}}.
\end{equation}

On the other hand, to effectively fuse the both extracted feature maps, i.e., $\mathbf{G}$ from the ConvNet and $\mathbf{S}$ from the AttenNet, we first concatenate them on the image-channel dimension for alignment and then adopt a concise CNN with an $a_f \times a_f \times2$-dimension convolutional kernel, referred to the FNet in Fig. 5(a), to fuse the both feature maps. Denote the output of FNet as $\mathbf{J}\in \mathbb{R} ^{N_\mathrm{s}\times N_\mathrm{t}\times 2}$, which is given by
\begin{equation}
\mathbf{J}=\mathbf{Conv2d}\left( \mathbf{Concat}\left( \mathbf{G},\mathbf{S} \right) ,\mathbf{\Theta }_{\mathrm{F}},\boldsymbol{\epsilon }_{\mathrm{F}},a_f \times a_f \times2 \right),
\end{equation}
where $\mathbf{Concat}$ denotes the concatenation operation on image-channel dimension and $\mathbf{\Theta }_{\mathrm{F}}\in \mathbb{R} ^{a_f \times a_f \times2}$ and $\boldsymbol{\epsilon }_{\mathrm{F}}\in \mathbb{R} ^{a_f \times2}$ are the weights and bias, respectively. 

\begin{figure}[t] 
\centering
    \includegraphics[scale=0.5]{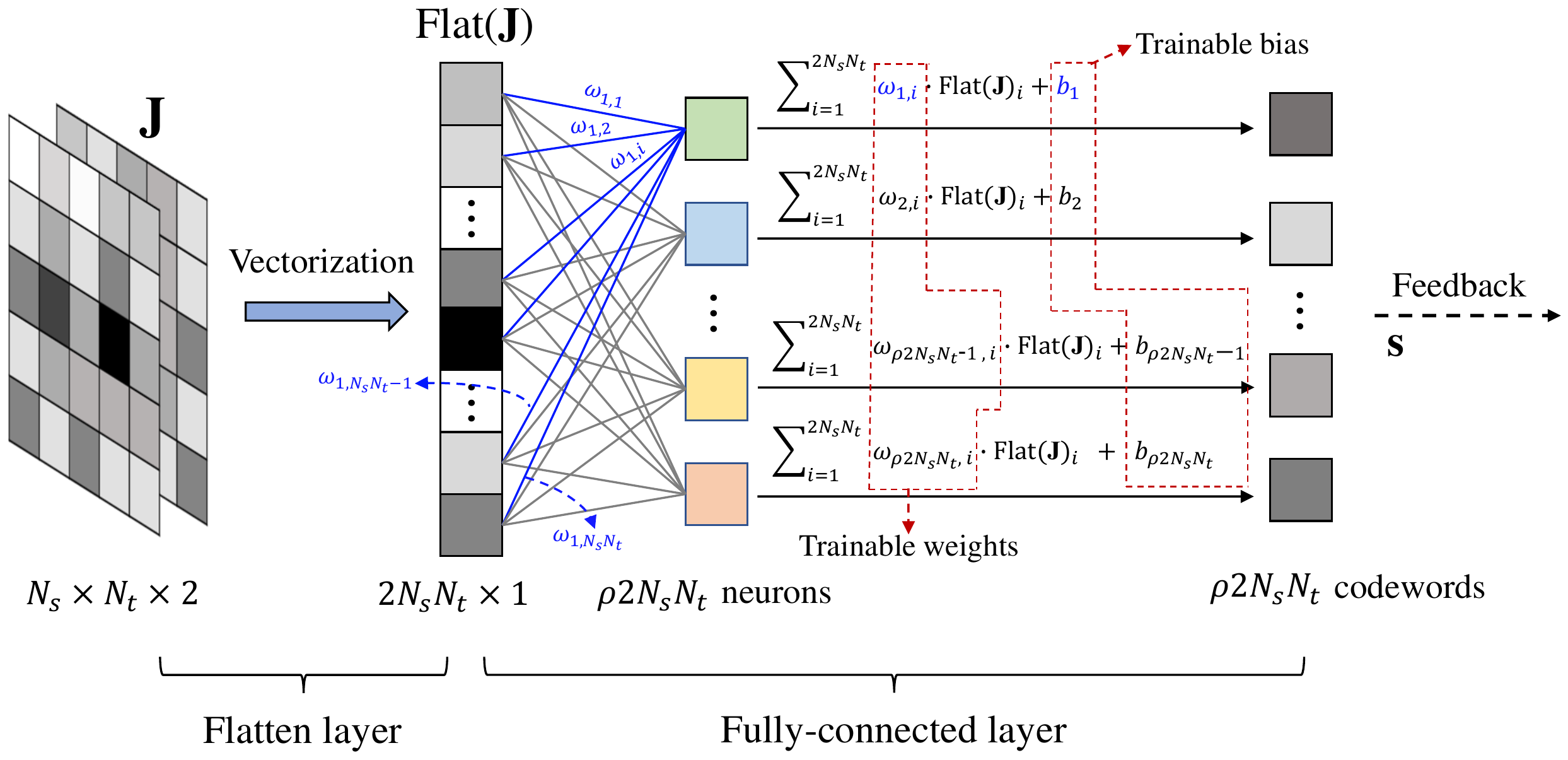}
    \caption{A diagram of the compression process in ComNet.}
\end{figure}

Following the DuffinNet, the final ComNet is used to compress the feature maps $\mathbf{J}$ into a codeword vector according to the compression ratio $\rho$. The ComNet is composed of a flatten layer and a fully-connected layer, as shown in the black dotted area of Fig. 5(a). For better understanding, we provide a diagram of the process of the proposed ComNet, as presented in Fig. 6. Specifically, we first transform $\mathbf{J}$ into a vector by a flatten layer, i.e., convert the dimension of $\mathbf{J}$ from ${N_\mathrm{s}\times N_\mathrm{t}\times 2}$ to a vector with ${2N_\mathrm{s}N_\mathrm{t}\times1}$ elements and then reduce its dimensionality by a fully-connected layer according to $\rho$. The final codeword vector is denoted as $\mathbf{s}\in \mathbb{R} ^{\rho 2N_\mathrm{s} N_\mathrm{t}\times 1}$ and the $j$-th element of $\mathbf{s}$ is expressed as
\begin{equation}
    s_j=\sum\limits_{i=1}^{2N_sN_t}{\omega _{j,i}}\cdot \mathrm{Flat}\left( \mathbf{J} \right) _i+b_j,\forall j\ =\ 1,2,\cdots, \rho 2N_sN_t,
\end{equation}
where the $\mathrm{Flat}\left( \mathbf{J} \right) _i \in \mathbb{R}$ denotes the $i$-th element of the flatten vector, $\omega _{j,i} \in \mathbb{R}$ represents the trainable weight between the $\mathrm{Flat}\left( \mathbf{J} \right) _i$ and the $j$-th neuron of the fully-connected layer, and $b_j \in \mathbb{R}$ denotes the trainable bias of the $j$-th neuron of the fully-connected layer. All the weights and bias are adaptively adjusted by the Duffin-CsiNet in the training. Finally, the encoder feeds back the codeword $\mathbf{s}$ to the BS through the feedback link.

\subsection{Decoder of Duffin-CsiNet}
The decoder network is used to reconstruct the CSI from the received codeword, i.e., $\mathbf{s}$. The detailed structure of the decoder network is shown in Fig. 5(b). It consists of three parts: the preprocessing network (PreNet), the cascaded DuffinNet, and the reconstruction network (RecNet). 

For the received codeword $\mathbf{s}$, the PreNet is the first processor in the decoder of Duffin-CsiNet which is used to transform the vector $\mathbf{s}$ into a $T$-image-channel feature map. It consists of a fully-connected layer, a reshape layer, and a composite convolutional layer. Specifically, the dimension of the codeword $\mathbf{s}$ is first raised from ${\rho 2N_\mathrm{s}N_\mathrm{t}\times 1}$ to ${2N_\mathrm{s}N_\mathrm{t}\times 1}$ by a fully-connected layer, then it is converted into an unprocessed image with a dimension of ${N_\mathrm{s}\times N_\mathrm{t}\times 2}$, denoted by $\mathbf{U}$, by a reshape layer, and finally we transform $\mathbf{U}$ into $T$ feature maps, denoted by $\mathbf{U}_\mathrm{e} \in \mathbb{R} ^{N_\mathrm{s}\times N_\mathrm{t}\times T}$, by the composite convolutional layer, which is represented by
\begin{equation}
    \mathbf{U}_\mathrm{e}=\mathbf{ComConv}_1\left( \mathbf{U},t_{\mathrm{e}}^{1}\times t_{\mathrm{e}}^{2}\times T \right),
\end{equation}
where $\mathbf{ComConv}_1$ denotes one composite convolutional layer and $t_{\mathrm{e}}^{1}\times t_{\mathrm{e}}^{2} \times T$ is the dimension of the convolutional kernel. 

For the procession of $\mathbf{U}_\mathrm{e}$, we exploit the cascaded OCN-based DuffinNet to extract its inherent feature. The OCN-based DuffinNet consists of three components: an $L$-layer ConvNet, an OCN-based AttenNet, and an FNet. In particular, the $L$-layer ConvNet and the FNet are set the same as that of the encoder network. Different from the AttenNet of the encoder, the AttenNet of the decoder applies the adaptive convolutional NN, i.e., OCN with an adaptive-size convolutional kernel, since OCN processes high-channel-dimension images more efficiently than the FNN of AttentNet of the encoder \cite{9156697}. 
% \begin{figure}[t]
%     \centering
%     \includegraphics[scale=0.45]{Decoder结构.pdf}
%     \caption{The structure of the decoder in Duffin-CsiNet.}
% \end{figure}

The OCN-based AttenNet is used to extract the dominant propagation-path features of the feature maps of $\mathbf{U}_\mathrm{e}$ that is the output of PreNet. As shown in the blue dotted box in Fig. 5(b), it consists of an average pooling layer and an adaptive convolutional layer. Specifically, $\mathbf{U}_\mathrm{e}$ is first processed by average pooling layer, which is expressed by
\begin{equation}
    \mathbf{u}_{\text{e}}=\mathbf{AveragePool}\left( \mathbf{U}_{\text{e}} \right),
\end{equation}
where $\mathbf{u}_{\text{e}} \in \mathbb{R} ^{1\times 1\times T}$ denotes the average value-based vector of $\mathbf{U}_{\text{e}}$ and $\mathbf{AveragePool}$ denotes the average pooling operator in (9). Then, an 1D convolutional layer which adopts adaptive-size convolution kernel is used to handle $\mathbf{U}_{\text{e}}$ to obtain the attention vector. Denote the size of the one-dimension convolutional kernel by $\mathrm{k}_\mathrm{adap}$, and the attention vector, denoted by $\mathbf{a}_{\text{u}} \in \mathbb{R} ^{1\times 1\times T}$, is given by \cite{9156697}
\begin{align}
    % \mathrm{k}_\mathrm{adap} & =2\lceil \frac{\lfloor \frac{\log _2T + 1}{2} \rfloor +1}{2} \rceil -1, \\
     \mathrm{k}_\mathrm{adap}  & = | \frac{\log _2T+1}{2} |_\mathrm{odd}, \\
    \mathbf{a}_{\mathrm{u}}  & =\mathbf{Conv1d}\left( \mathbf{u}_{\mathrm{e}},\mathrm{k}_\mathrm{adap} \right),
\end{align}
where $| x |_\mathrm{odd}$ represents the nearest odd integer to $x$ and $\mathbf{Conv1d}$ denotes the operation of 1D convolution. Applying the attention mechanism, the final dominant propagation-path feature maps, denoted as $\mathbf{U}_{\mathrm{d}}\in \mathbb{R} ^{N_\mathrm{s}\times N_\mathrm{t}\times T}$, is obtained by
\begin{equation}
    \mathbf{U}_{\mathrm{d}}=\mathrm{\sigma}\left( \mathbf{a}_{\mathrm{u}} \right) \otimes \mathbf{U}_{\mathrm{e}},
\end{equation}
where $\mathrm{\sigma}(\cdot)$ denotes the Sigmoid function defined in (13).

% \begin{figure*}[t]
%     \centering
%     \includegraphics[scale=0.453]{Duffin-CsiNet2.jpg}
%     \caption{The overall architectures of Duffin-CsiNet.}
% \end{figure*}

\begin{figure}[t]
    \centering
    \includegraphics[scale=0.55]{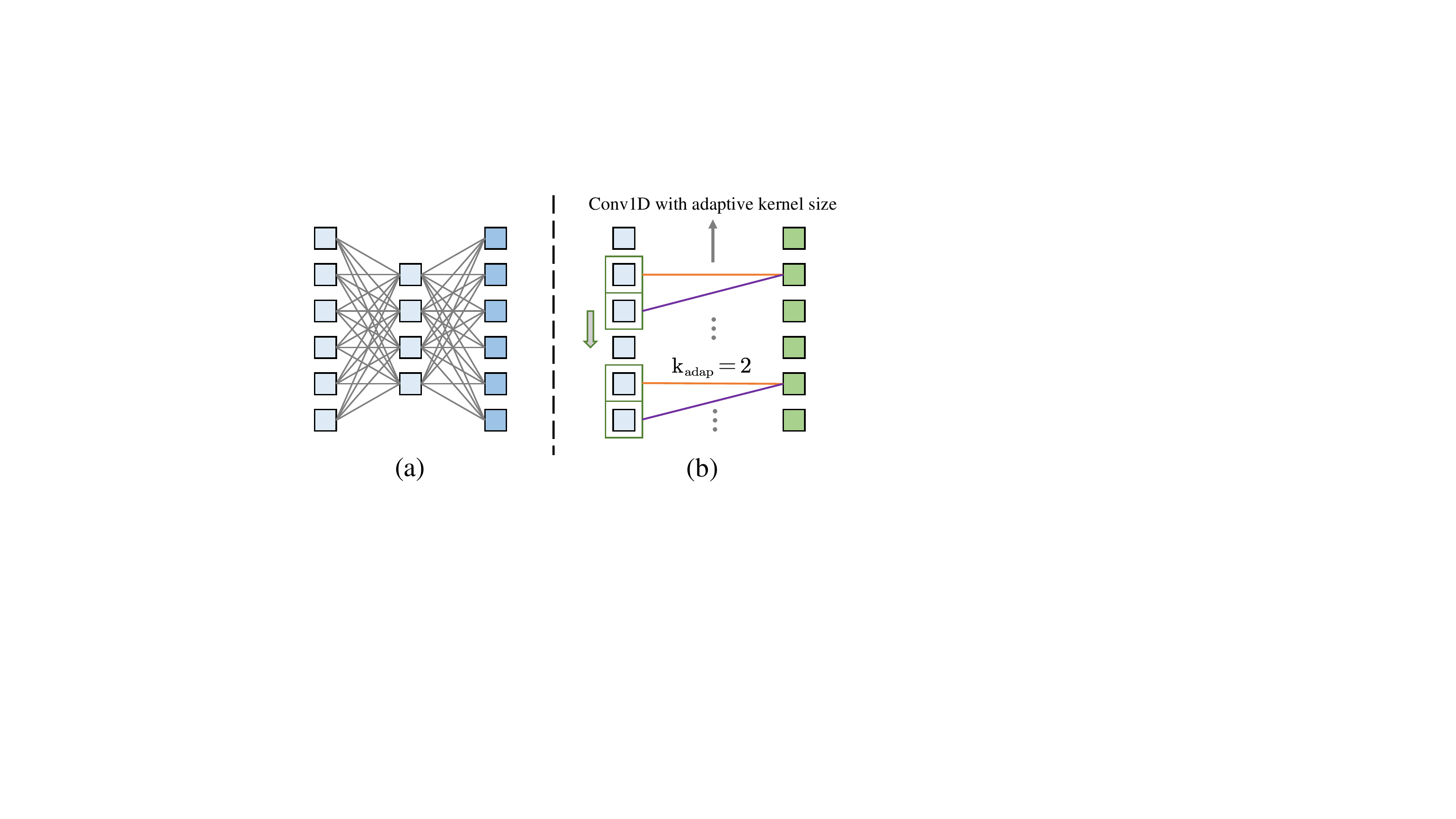}
    \caption{(a) Fully-connected layer; (b) Adaptive convolutional layer.}
\end{figure}

In Fig. 7, we further illustrate the structures of a fully-connected layer and an adaptive convolutional layer. In particular, the fully-connected layer connects each neural node, which is suitable for CSI feature extraction and compression as it considers the correlation among all the elements of the original CSI when extracting the physical features of CSI image. In contrast, the adaptive convolution layer exploit one-dimension convolution and connects each the neural node to its adjacent nodes while without altering the dimension, which is usually used for the feature extraction of high-channel-dimension feature image to improve network performance \cite{9156697}.

Moreover, to effectively extract the inherent features of the unprocessed image $\mathbf{U}_{\text{e}}$, we apply the structure of cascaded DuffinNet. In particular, as shown in Fig. 5(b), we deploy multiple DuffinNets in the decoder and the parameter setting of each DuffinNet is identical.

% \begin{figure}[t]
% \subfigure[Fully-connected Layer]{
% \begin{minipage}[t]{0.5\linewidth}
% \centering
% \includegraphics[scale=0.6]{FNN.pdf}
% \end{minipage}%
% }% awq
% \subfigure[Adaptive convolutional Layer]{
% \begin{minipage}[t]{0.5\linewidth}
% \centering
% \includegraphics[scale=0.6]{ECA.pdf}
% %\caption{fig1}
% \end{minipage}%
% }%
% \caption{The connected structures of the fully-connected layer and the adaptive convolution layer.}
% % \centerline{\ \ \small{ Fig. 4.}}
% \end{figure}

Following the cascaded DuffinNet, a RecNet composed of a composite convolutonal layer is used to recover the CSI image from the feature maps $\mathbf{U}_{\mathrm{d}}$, which is expressed by
\begin{equation}
\hat{\mathbf{{H}}}_\mathrm{s}=\mathbf{ComConv}_1\left( \mathbf{U}_{\mathrm{d}},t_{\mathrm{d}}^{1}\times t_{\mathrm{d}}^{2}\times 2 \right),
\end{equation}
where $t_{\mathrm{d}}^{1}\times t_{\mathrm{d}}^{2}\times 2$ is the dimension of convolutional kernel. In particular, the activation function in (24) is Sigmoid function instead of LeakyReLU function.
% Finally, the CSI reconstruction image $\mathbf{\hat{H}}_s$ is obtained by applying the sigmoid activation function on $\mathbf{U}_{\mathrm{dr}}$.

\subsection{Training and Deployment}

For the training of Duffin-CsiNet, we adopt the approach of supervised learning. In particular, the loss function of mean squared error (MSE) is exploited for back propagation of the overall network, which is expressed as
\begin{equation}
    \mathrm{Loss}=\frac{1}{T_\mathrm{s}}\sum_{i=1}^{T_\mathrm{s}}{\left\| \mathbf{\hat{H}}_{\mathrm{s}}\left[ i \right] -\mathbf{H}_{\mathrm{s}}\left[ i \right]  \right\| ^2},
\end{equation}
where $T_\mathrm{s}$ represents the total number of training samples $\mathbf{H}_\mathrm{s}$. The Duffin-CsiNet carries out gradient update and improves the effect of CSI reconstruction by minimizing the value of the $\mathrm{Loss}$ in (25). Besides, we use the NMSE to evaluate the performance of CSI reconstruction of Duffin-CsiNet, which is defined as
\begin{equation}
    \mathrm{NMSE}=\mathbb{E} \left\{ \frac{\left\| \mathbf{\hat{H}}_\mathrm{s}-\mathbf{H}_\mathrm{s} \right\| ^2}{\left\| \mathbf{H}_\mathrm{s} \right\| ^2} \right\}.
\end{equation}
Compared to MSE, the NMSE can fairly capture the performance gap of different methods, which has been widely used in existing studies, e.g., in \cite{8543184, 9445070, lu2021binarized, 9419066, yzq}.
% Note that $\mathrm{NMSE}$ is generally a negative number
% , and in the following simulation analysis, sometimes we also use the absolute value of $\mathrm{NMSE}$, i.e., $\left| \mathrm{NMSE} \right|$, as an evaluation indicator.

\begin{figure*}[t]
    \centering
    \includegraphics[scale=0.45]{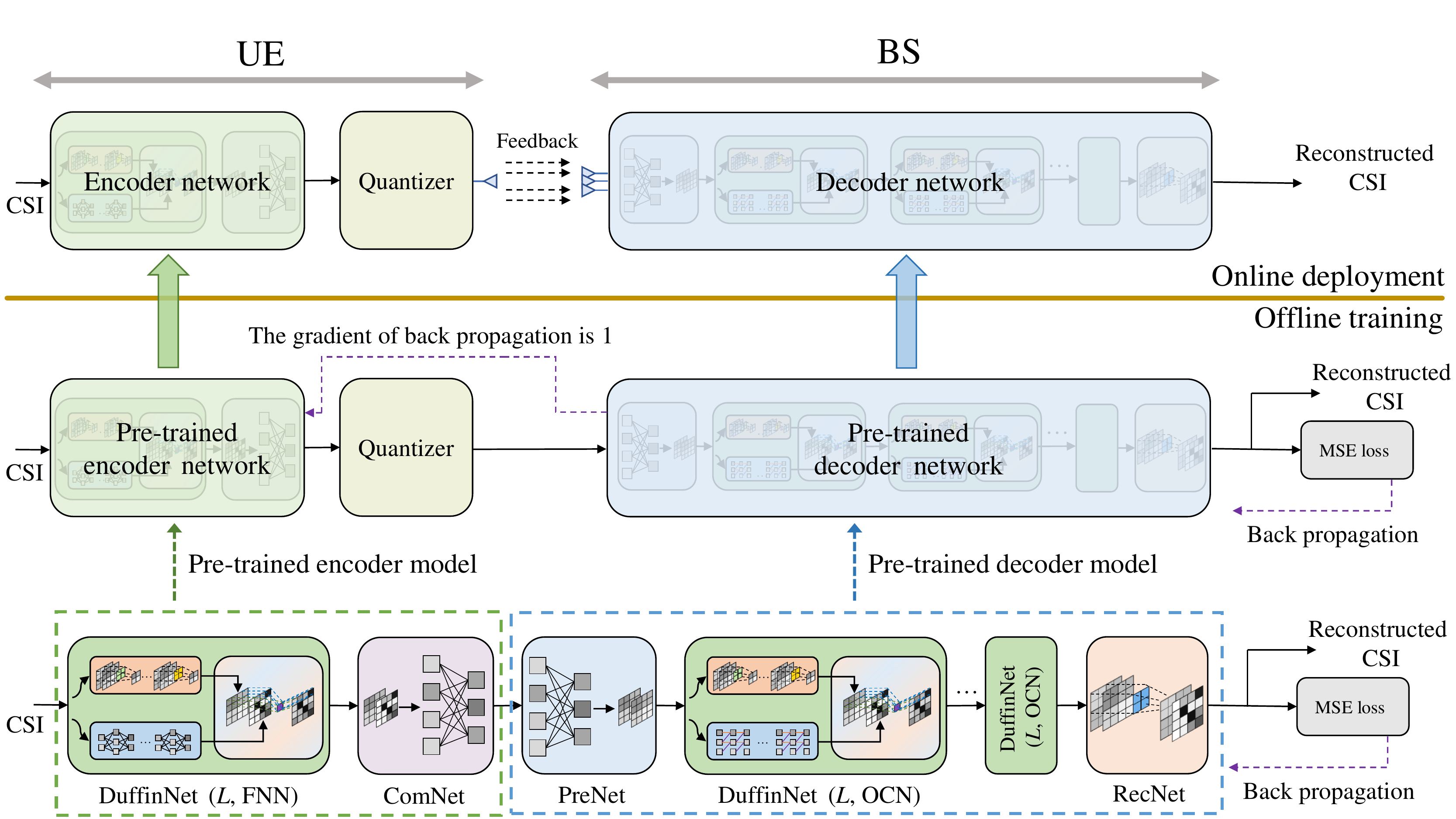}
    \caption{The diagram of the training and deployment of the proposed Duffin-CsiNet with quantization.}
\end{figure*}

Moreover, we exploit the strategy of ``warm up" \cite{9149229} to adapt the learning rate in the training stage, which is expressed as 
\begin{equation}
    \iota =\iota _\mathrm{min}+\frac{1}{2}\left( \iota _\mathrm{max}-\iota _\mathrm{min} \right) \left( 1+\cos \left( \frac{t_\mathrm{s}-T_\mathrm{w}}{T_\mathrm{e}-T_\mathrm{w}}\pi \right) \right),
\end{equation}
where $\iota$, $\iota _\mathrm{min}$, and $\iota _\mathrm{max}$ denote the current, initial, and final learning rate, respectively. $t_\mathrm{s}$ is the value of current epoch, $T_\mathrm{w}$ denotes the number of ``warm up" epoch, and $T_\mathrm{e}$ is the total number of training epoch. Different from other existing popular ways, e.g., fixing the learning rate \cite{8322184} and proportionally decreasing the learning rate \cite{9439959}, the ``warm up" strategy first lets the NNs quickly learn by increasing the learning rate in the early epoch and then slow down the speed of gradient descent by gradually reducing the learning rate, which can in turn speed up the overall the network convergence and improve the learning performance \cite{9149229}.

For practical applications of the proposed Duffin-CsiNet, to address the challenges of transmitting continuous codeword values \cite{9347820}, we need to quantize the compressed codeword $\mathbf{s}$. Existing DL-based quantization designs, e.g., \cite{8845636,9296555,9090892,9799802,8461983,9057584}, have shown that applying quantization to network training is an effective way to improve the performance of low-resolution quantization. Following this idea, we adopt a two-stage approach with quantization, i.e., pre-training-based offline training with quantization and online deployment with quantization. As shown in Fig. 8, we first train the proposed Duffin-CsiNet without quantizing the feedback codewords for a satisfactory pre-trained network model. After the pre-training is completed, we deploy the uniform quantizer as in \cite{8845636,9296555,9090892,9799802} after the encoder network and then retrain the pre-trained Duffin-CsiNet. As the derivative of the quantization function is zero almost everywhere, it cannot be applied in the back propagation of Duffin-CsiNet for gradient update \cite{8845636}. In order to address this problem, we adopt the quantization method with gradient forgery strategy applied in \cite{8845636} to mimic the gradient of quantization during the back propagation of Duffin-CsiNet in the stage of training. In particular, the quantization gradient in the training of Duffin-CsiNet is always set to 1.

% Specifically, the stage of offline training is carried out without quantization operation at the BS. When the offline training is completed, the encoder network is deployed at the UE and the decoder network is deployed at the BS, and the Lloyd-Max quantizer \cite{1056489} is exploited to accomplish the quantization. As for the online using stage, the codeword values from the trained encoder network are quantized by the Lloyd-Max quantizer and then fed back to the decoder at the BS for recovering the CSI through the corresponding dequantizer and the trained decoder. 

\begin{figure*}[t]
    \centering
    \includegraphics[scale=0.4]{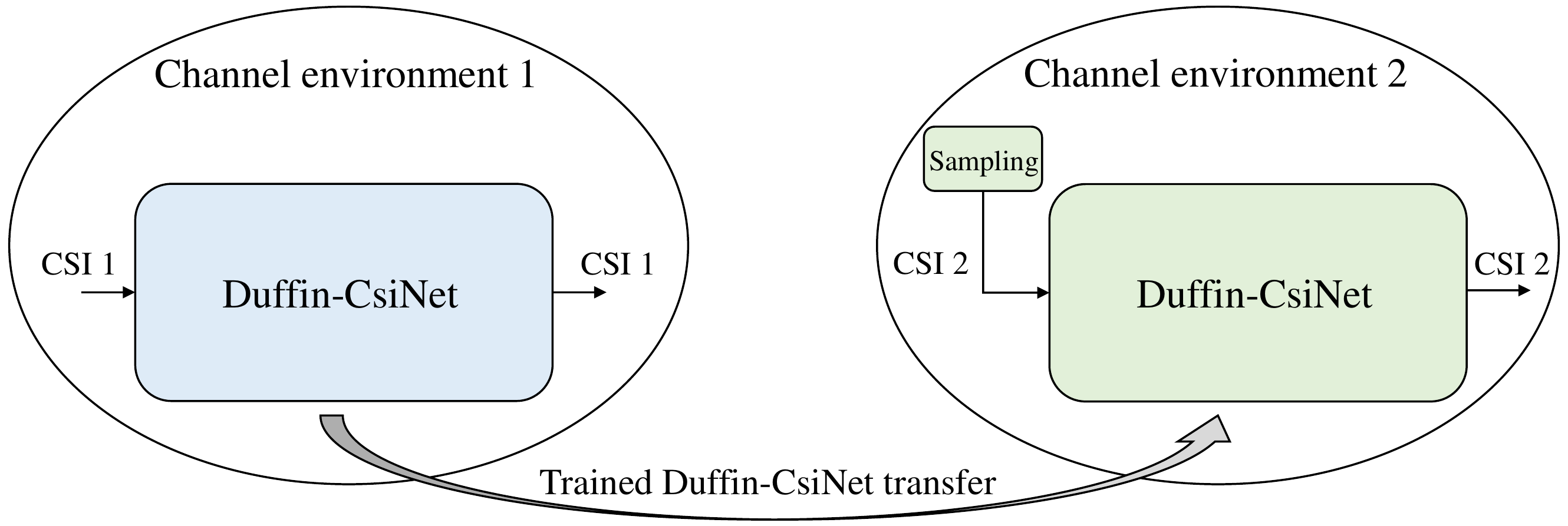}
    \vspace{-1em}
    \caption{Transfer learning-based strategy.}
\end{figure*}

On the other hand, for improving the generalization of Duffin-CsiNet, we propose a transfer learning-based strategy, which allows the NNs to quickly adapt to varying channel environments. The diagram of the proposed transfer learning-based strategy is shown in Fig. 9. Specifically, we have a trained Duffin-CsiNet under a channel environment and it enjoys an excellent CSI reconstruction performance. When the channel environment changes, we can directly transfer the trained Duffin-CsiNet to better fit the current environment. We only need to sample some CSI under the current channel environment to refine the trained Duffin-CsiNet for improving the performance. This transfer learning approach effectively utilizes existing NN knowledge and achieves satisfactory performance with less training than a full retraining.

It is worth noting that after retraining the Duffin-CsiNet at the BS by adopting the transfer learning-based strategy, the parameters of the encoder network need to be exchanged with the UE, which may bring excessive wireless resource overhead, especially when the channel environment varies frequently. To address the tremendous wireless resource overhead caused by the updating of encoder network, designing an one-side-NN CSI feedback architecture serves as a promising solution. Specifically, a trainable NN is only deployed in the decoder at the BS, while a feature extraction algorithm requiring no training is exploited for CSI compression at the UE. This one-side framework saves the updating overhead of encoder network parameters. Note that it is challenging to design effective feature extraction algorithms and the corresponding decoupling algorithms, providing an interesting direction for future research.

% for the extension to the system where multiple UEs exist, we can directly apply the well-trained Duffin-CsiNet framework in each UE for CSI compression. Further considering the channel difference between the BS and different UEs, we can also adopt some effective strategies, e.g., first deploy the well-trained Duffin-CsiNet at each UE, and then fine-tune the corresponding Duffin-CsiNet through less current CSI sample, which is also an interesting direction well worth studying in the future.

\section{Simulations}

In this section, we illustrate the performance of the proposed Duffin-CsiNet. We first describe the datasets and the network setting. Then, we consider the following aspects to verify the effectiveness of the proposed Duffin-CsiNet: NMSE performance, visualization of CSI feature extraction, quantization, generalization, network complexity and convergence, bit error rate (BER) for the reconstructed CSI, and the ablation experiments.

% $\bullet$ NMSE comparison of CSI reconstruction with quantization.

% $\bullet$ Visualization of CSI compression.

% $\bullet$ Generalization of networks.

% $\bullet$ Discussion on network complexity, convergence.

% $\bullet$ System performance analysis for reconstructed CSI.

% including NMSE performance with and without quantization, network complexity, network convergence, discussion for the influence of $T$, bit error rate (BER) for reconstructed CSI.
% \begin{itemize}
% \item NMSE performance with and without quantization.
% \item Network complexity.
% \item Network convergence.
% \item Discussion for the influence of $T$.
% \item Sum rate for reconstructed CSI.
% \end{itemize}

\subsection{Simulation Setup}

% % Table generated by Excel2LaTeX from sheet 'Sheet1'

\subsubsection{Datasets}
As many previous works \cite{9445070, 8972904, 9171358, 9439959, lu2021binarized, 9497358, 9419066}, we follow the experimental settings used in CsiNet \cite{8322184}. The training sets and the testing sets are generated according to the COST 2100 channel model \cite{6393523} and we consider two scenarios: the indoor picocellular scenario at the 5.3 GHz frequency band and the outdoor rural scenario at the 300 MHz frequency band. An FDD system with $N_\mathrm{c} = 1024$ subcarries is adopted. For the considered massive MIMO system, the BS is equipped with a uniform linear array with $N_\mathrm{t} = 32$ antennas. After transforming the original CSI matrix into the angular-delay domain as in (2), the first $N_\mathrm{s} = 32$ rows of $ \mathbf{H}_\mathrm{d}$ in the delay domain are retained. The training set and the testing set contain 100,000 and 20,000 samples of $ \mathbf{H}_\mathrm{s}$, respectively. We adopt the training sets during the offline training process. When the training process is completed and the trained network model is obtained, we apply it to the test by exploiting the testing set, which corresponds to the online stage.

\begin{table}[t]
\caption{Setting of the proposed Duffin-CsiNet}
\begin{minipage}{0.5\linewidth}
\centering
% \caption{NMSE (dB) of the NNs in indoor scenario}
\vspace{-1.5em}
    \scalebox{0.5}{
    \begin{tabular}{c|c|c|p{20em}}
    \toprule
    \multicolumn{1}{c}{} & \textbf{Sub-network} & \textbf{Layer name} & \multicolumn{1}{c}{\textbf{Parameter}} \\
    \midrule
    \multirow{15}[8]{*}{\textbf{Encoder}} & \multirow{7}[2]{*}{AttenNet} & Average Pooling $($Max Pooling$)$ & \multicolumn{1}{c}{N/A} \\
          &       & Fully-connected  & \multicolumn{1}{c}{1 neuron} \\
          &       & Batch normalization & \multicolumn{1}{c}{$\zeta = 1 \times 10^{-5}$} \\
          &       & Activation & \multicolumn{1}{c}{ReLU} \\
          &       & Fully-connected  & \multicolumn{1}{c}{2 neurons} \\
          &       & Batch normalization & \multicolumn{1}{c}{$\zeta = 1 \times10^{-5}$} \\
          &       & Activation & \multicolumn{1}{c}{Sigmoid} \\
\cmidrule{2-4}          & \multirow{5}[1]{*}{ConvNet} & Composite 2D convolution & \makecell{2 kernels of 3 $\times$ 3, stride 1, padding (1, 1), \\ $\zeta$ = 1$\times 10^{-5}$, LeakyReLU, $\alpha$ = 0.3}  \\
          &       & Composite 2D convolution & \makecell{2 kernels of 1 $\times$ 9, stride 1, padding (0, 4), \\ $\zeta$ = 1$\times 10^{-5}$, LeakyReLU, $\alpha$ = 0.3}  \\
          &       & Composite 2D convolution & \makecell{2 kernels of 9 $\times$ 1, stride 1, padding (4, 0), \\ $\zeta$ = 1$\times 10^{-5}$, LeakyReLU, $\alpha$ = 0.3}  \\
\cmidrule{2-4}          & FNet & 2D convolution & \multicolumn{1}{c}{2 kernels of 3 $\times$ 3, stride 1, padding (0, 0)} \\
\cmidrule{2-4}          & \multirow{2}[1]{*}{ComNet} & Flatten & \multicolumn{1}{c}{N/A} \\
          &       & Fully-connected  & \multicolumn{1}{c}{2048$\times \rho$ neurons} \\
    \bottomrule
    \end{tabular}} %
\end{minipage}
\begin{minipage}{0.5\linewidth}  
\centering
% \caption{NMSE (dB) of the NNs in outdoor scenario}
\vspace{-1.5em}
    \scalebox{0.49}{
    \begin{tabular}{c|c|c|p{20em}}
    \toprule
    \multicolumn{1}{c}{} & \textbf{Sub-network} & \textbf{Layer name} & \multicolumn{1}{c}{\textbf{Parameter}} \\
    \midrule
    \multirow{14}[10]{*}{\textbf{Decoder}} & \multirow{3}[1]{*}{PreNet} & Fully-connected & \multicolumn{1}{c}{2048 neurons} \\
          &       & Reshape & \multicolumn{1}{c}{N/A} \\
          &       & Composite 2D convolution & \makecell{64 kernels of 5 $\times$ 5, stride 1, padding (2, 2), \\$\zeta$ = 1$\times 10^{-5}$, LeakyReLU, $\alpha$ = 0.3} \\
\cmidrule{2-4}          & \multirow{3}[1]{*}{AttenNet} & AveragePool & \multicolumn{1}{c}{N/A} \\
          &       & 1D convolution & \multicolumn{1}{c}{1 kernel of 3 $\times$ 1, stride 1, padding $(2, 0)$} \\
          &       & Activation & \multicolumn{1}{c}{LeakyReLU, $\alpha$ = 0.3 } \\
\cmidrule{2-4}          & \multirow{5}[1]{*}{ConvNet} & Composite 2D convolution & \makecell{64 kernels of 3 $\times$ 3, stride 1, padding (1, 1), \\ $\zeta$ = 1$\times 10^{-5}$, LeakyReLU, $\alpha$ = 0.3}  \\
          &       & Composite 2D convolution & \makecell{64 kernels of 1 $\times$ 9, stride 1, padding (0, 4), \\$\zeta$ = 1$\times 10^{-5}$, LeakyReLU, $\alpha$ = 0.3}  \\
          &       & Composite 2D convolution & \makecell{64 kernels of 9 $\times$ 1, stride 1, padding (4, 0), \\ $\zeta$ = 1$\times 10^{-5}$, LeakyReLU, $\alpha$ = 0.3}  \\
\cmidrule{2-4}          & FNet & 2D convolution & \multicolumn{1}{c}{2 kernels of 3 $\times$ 3, stride 1, padding (0, 0)} \\
\cmidrule{2-4}          & RecNet & Composite 2D convolution & \makecell{2 kernels of 5 $\times$ 5, stride 1, padding (2, 2), \\ $\zeta$ = 1$\times10^{-5}$, Sigmoid} \\
    \bottomrule
    \end{tabular}} %%
\end{minipage}
\end{table}

\subsubsection{Networks setting and simulation device} The parameter setting of Duffin-CsiNet is summarized in Table I. In particular, the two FNNs in the AttenNet of the encoder are identical. Moreover, $T$ is selected as 64 and $\mathrm{k}_\mathrm{adap} = $ 3 is obtained by substituting the corresponding $T$ into (21). The $\omega _\mathrm{a}$ and $\omega _\mathrm{v}$ in the encoder are set to 1.0 and 0.5 at the beginning of network training, respectively, and the convolutional kernel size $a_f$ is set to 3 in the FNet. The number of DuffinNet in the decoder network is selected as 2. We adopt the end-to-end supervised training. The weights and bias of all the convolutional layers and fully-connected layers are initialized randomly and the Adam optimizer \cite{lu2021binarized, 9497358, 9419066} is used to update the training parameters by the back propagation of NNs. The number of training epoch $T_\mathrm{e}$ is set to be 1500, the batch size is 200, and the learning rate exploits ``warm up" strategy and we set the parameters of (27) to be same as that of \cite{9149229} for fair comparison, i.e., $\iota _\mathrm{min} = 5\times10^{-5}$, $\iota _\mathrm{max} = 2\times10^{-3}$, and $T_\mathrm{w} = 30$. 
% With the help of warm up-aided cosine annealing scheduler, the Adam optimizer can significantly improve the network learning performance \cite{lu2021binarized}. 
All subsequent simulations, including the re-experiments of the compared methods, are carried out with the same training settings for a fair comparison in Python 3.8.8 with Pytorch 1.8.0 on an NVIDIA GTX3090 GPU.

\subsection{NMSE Performance}
% % encoder的flops
% \begin{table*}[h]
%   \centering
%   \caption{Encoder FLOPs comparison}
%   \scalebox{0.7}{
%     \begin{tabular}{c|c|c|c|c}
%     \toprule
%     Method & $\rho$ = 1/4 & $\rho$ = 1/16 & $\rho$ = 1/32 & $\rho$ = 1/64 \\
%     \midrule
%     CLNet \cite{9497358} & 1.434 & 0.648 & 0.517 & 0.451 \\
%     ACRNet \cite{lu2021binarized} & 1.319 (- 8.0\%) & 0.533 (- 17.7\%) & 0.402 (- 22.2\%) & 0.336 (- 25.5\%) \\
%     \rowcolor[rgb]{ .906,  .902,  .902} Duffin-CsiNet\_Base & \textbf{1.196 (- 16.6\%)} & \textbf{0.409 (- 36,9\%)} & \textbf{0.278 (- 46.2\%)} & \textbf{0.213 (- 52.8\%)} \\
%     \rowcolor[rgb]{ .906,  .902,  .902} Duffin-CsiNet\_Midd & \textbf{1.196 (- 16.6\%)} & \textbf{0.409 (- 36,9\%)} & \textbf{0.278 (- 46.2\%)} & \textbf{0.213 (- 52.8\%)} \\
%     \rowcolor[rgb]{ .906,  .902,  .902} Duffin-CsiNet\_Large & \textbf{1.196 (- 16.6\%)} & \textbf{0.409 (- 36,9\%)} & \textbf{0.278 (- 46.2\%)} & \textbf{0.213 (- 52.8\%)} \\
%     \bottomrule
%     \end{tabular}}%
% \end{table*}%

To validate the performance of the proposed Duffin-CsiNet for CSI compression and reconstruction, we compare the its NMSE with existing DL-based CSI compression methods. In particular, we select four representative methods using the architecture of CNNs, including CsiNet \cite{8322184}, BCsiNet \cite{9373670}, CRNet \cite{9149229}, ACRNet \cite{lu2021binarized}, and a SOTA method using the architecture of ANNs, i.e., CLNet \cite{9497358}. Specifically, we perform the experimental simulations under five different compression ratios, including 1/4, 1/8, 1/16, 1/32, and 1/64, and apply them in two channel environment scenarios. 

The NMSE results of indoor scenarios are shown in Table II. It can be observed that the proposed Duffin-CsiNet outperforms the others five existing methods in terms of the NMSE performance in all the considered compression ratios. In particular, when the compression ratio is $\rho$ = 1/4, the NMSE results of the proposed Duffin-CsiNet is -35.19 dB, which has about 6 dB gain compared to the CLNet, about 5.5 dB gain compared to the ACRNet, and about 21 dB gain compared to the classic CsiNet. As for the case of high compression ratio, e.g., $\rho$ = 1/32, the proposed Duffin-CsiNet still enjoys a superior NMSE performance compared to the existing methods, e.g., the Duffin-CsiNet has about 3 dB gain compared to the CsiNet and about 2 dB gain compared to ACRNet. The NMSE results of outdoor scenario are shown in Table III. Similar to the observation of indoor scenario, the proposed Duffin-CsiNet enjoys a satisfactory reconstruction performance.

\begin{table}[t]
\begin{minipage}{0.5\linewidth}
\centering
\caption{NMSE (dB) of the NNs in indoor scenario}
\vspace{-1.5em}
\scalebox{0.7}{
\begin{tabular}{c|c|c|c|c|c}
\hline
\multicolumn{1}{c|}{\diagbox{Method}{$\rho$}} & \multicolumn{1}{c|}{1/4} & \multicolumn{1}{c|}{1/8} & \multicolumn{1}{c|}{1/16} & \multicolumn{1}{c|}{1/32} & \multicolumn{1}{c}{1/64} \\
\hline
CsiNet \cite{8322184} & -13.97 & -12.14 & -9.69 & -8.62 & -5.36\\
\hline
BCsiNet \cite{9373670} & -24.19 & -12.38 & -10.42 & -9.11 & -6.64 \\
\hline
CRNet \cite{9149229} & -25.19 & -16.01 & -11.52 & -8.87 & -6.27 \\
\hline
CLNet \cite{9497358}& -29.04 & -15.60 & -11.05 & -8.63 & -6.25 \\
\hline
ACRNet \cite{lu2021binarized} & -29.83 & -18.73 & -13.34 & -9.62 & -7.74 \\
\hline
% \rowcolor[rgb]{ .906,  .902,  .902} Duffin-CsiNet\_Base & \textbf{-31.78} & x & \textbf{-15.41} & \textbf{-10.33} & \textbf{-7.95} \\
Proposed Duffin-CsiNet & \textbf{-35.19} & \textbf{-23.59} & \textbf{-17.20} & \textbf{-11.69} & \textbf{-8.05} \\
% \rowcolor[rgb]{ .906,  .902,  .902} Duffin-CsiNet\_Large & \textbf{-36.34} & x & \textbf{-17.38} & \textbf{-12.26} & \textbf{-9.02} \\
\hline
\end{tabular}}%
\end{minipage}
\begin{minipage}{0.5\linewidth}  
\centering
\caption{NMSE (dB) of the NNs in outdoor scenario}
\vspace{-1.5em}
\scalebox{0.7}{
\begin{tabular}{c|c|c|c|c|c}
\hline
\multicolumn{1}{c|}{\diagbox{Method}{$\rho$}} & \multicolumn{1}{c|}{1/4} & \multicolumn{1}{c|}{1/8} & \multicolumn{1}{c|}{1/16} & \multicolumn{1}{c|}{1/32} & \multicolumn{1}{c}{1/64} \\
\hline
CsiNet \cite{8322184} & -10.94 & -6.47 & -4.87 & -3.13 & -1.93\\\hline
BCsiNet \cite{9373670} & -11.80 & -6.26 & -4.98 & -3.35 & -2.16 \\\hline
CRNet \cite{9149229} & -12.32 & -8.04 & -5.41 & -3.51 & -2.22 \\\hline
CLNet \cite{9497358}& -12.88 & -8.29 & -5.56 & -3.49 & -2.19 \\\hline
ACRNet \cite{lu2021binarized} & -13.55 & -9.22 & -6.30 & -3.83 & -2.61 \\\hline
% \rowcolor[rgb]{ .906,  .902,  .902} Duffin-CsiNet\_Base & \textbf{-31.78} & x & \textbf{-15.41} & \textbf{-10.33} & \textbf{-7.95} \\
Proposed Duffin-CsiNet & \textbf{-16.12} & \textbf{-11.55} & \textbf{-7.84} & \textbf{-5.52} & \textbf{-3.85} \\
% \rowcolor[rgb]{ .906,  .902,  .902} Duffin-CsiNet\_Large & \textbf{-36.34} & x & \textbf{-17.38} & \textbf{-12.26} & \textbf{-9.02} \\
\hline
\end{tabular}}%
\end{minipage}
\end{table}
% In particular, when the compression ratio is $\rho$ = 1/4, the NMSE of the proposed Duffin-CsiNet achieves -16.12 dB, which has about 3 dB gain compared to the ACRNet and about 5 dB gain compared to the CsiNet. As for the case of high compression ratio, the proposed Duffin-CsiNet has a performance gain of at least 1 dB.

% On the other hand, from observing Table II, we find that the proposed Duffin-CsiNet only needs half of the compression ratio compared with other NNs when a specific NMSE value is needed, especially in the case of high compression ratio. For example, when the NMSE is required as -12 dB, the existing methods require a minimum compression ration of 1/16, while the Duffin-CsiNet can use compression ratio 1/32 to achieve this requirement. In other words, Duffin-CsiNet offers a huge potential in reducing the bandwidth overhead without performance degradation, providing a feasible solution for practical engineering. These present the effectiveness of the proposed Duffin-CsiNet for improving the CSI reconstruction, which results from the efficient parallel-serial hybrid structure that fully considers inherent physical features of CSI in the proposed DuffinNet. 

\begin{figure}[t]
    \centering
    \includegraphics[scale=0.45]{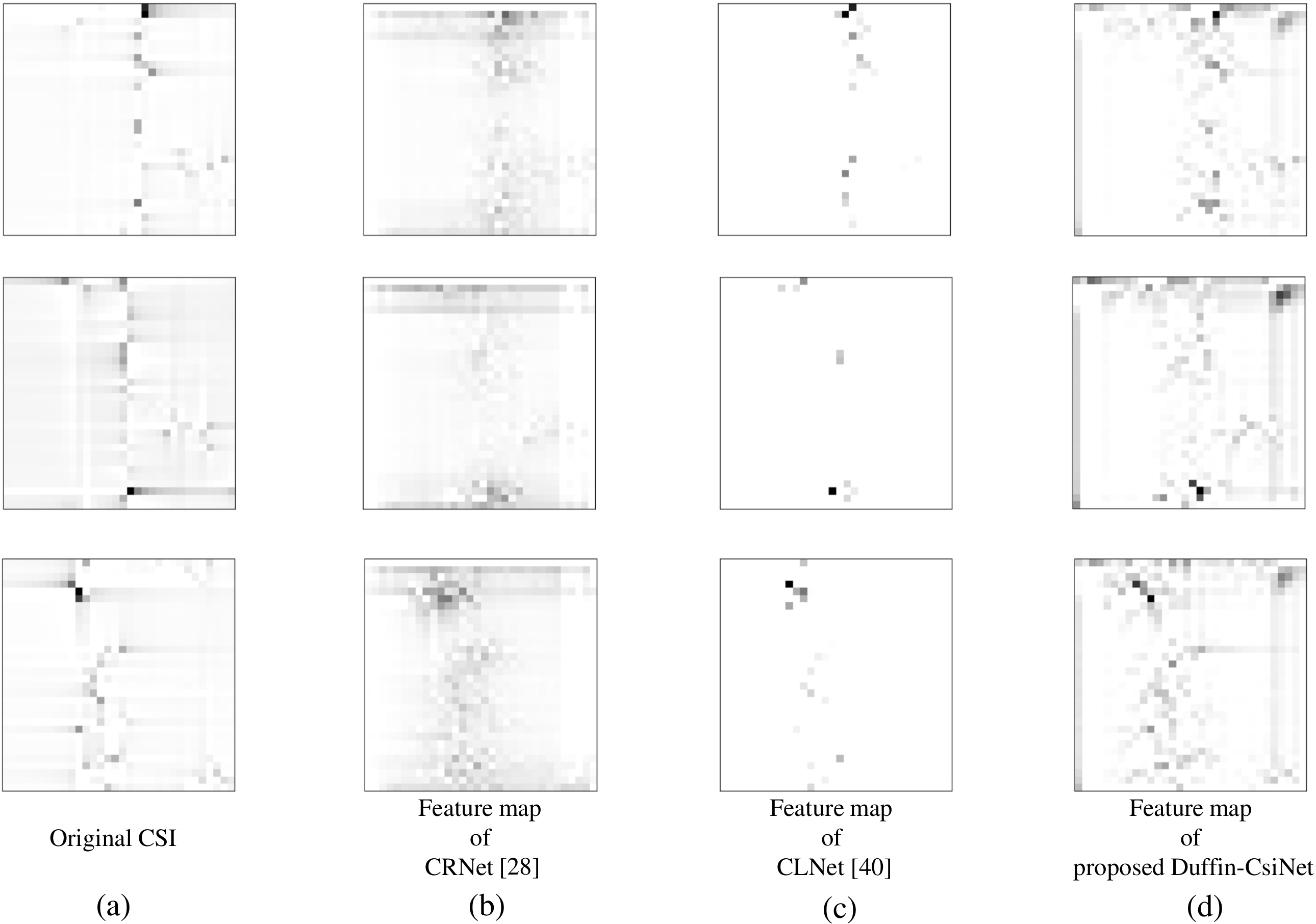}
    \vspace{-1.5em}
    \caption{Visualization comparison of CSI feature extraction. The (a) column is the original CSI image; (b) column, (c) column, and (d) column are the extracted feature maps of CRNet, CLNet and proposed Duffin-CsiNet, respectively.}
\end{figure}

\subsection{Visualization of CSI Feature Extraction}
To verify the effectiveness of the proposed dual-feature fusion enhanced method for CSI feature extraction, we present the encoded feature maps of Duffin-CsiNet versus the other networks, as shown in Fig. 10. Note that the encoded feature map refers to the feature map extracted by encoder before compressed into a codeword vector, i.e., $\mathbf{J}$ in (17). In particular, two outdoor channel images with rich scatters are selected for visualization.

From Fig. 10, we observe that the proposed Duffin-CsiNet performs dual-feature extraction on CSI images compared to the CNN-based CRNet \cite{9149229} and the ANN-based CLNet \cite{9497358}. It illustrates that Duffin-CsiNet outperforms the other two methods on the CSI feature extraction.
Specifically, the proposed Duffin-CsiNet not only extracts the NLOS propagation-path features, but greatly retains the dominant propagation-path features in the original CSI image, providing a comprehensive representative feature map for CSI compression compared to the existing NNs. The powerful feature extraction of DuffinNet leads to the improved CSI reconstruction and successfully solves the problem of limited feature extraction ability of the existing NNs.

% \begin{figure*}[t]
%     \centering
%     \includegraphics[scale=0.4]{量化结果对比放论文中.eps}
%     \caption{CSI reconstruction comparison at different quantization bits, $\rho = 1/8$.}
% \end{figure*}
\vspace{-1em}
\subsection{Quantization}
This section discusses the performance of the adopted two-stage training and deployment approach in Fig. 8. Taking $\rho$ = 1/8 as an example, we compare the performance of Duffin-CsiNet adopting quantization with several SOTA DL-based quantized feedback networks, including CsiQnet \cite{9090892}, DualQnet \cite{9090892}, JCNet \cite{8845636}, CH-CsiNetPro-PQB \cite{9799802}, and CH-DUalNetSph-PQB \cite{9799802}. The corresponding simulation results are shown in Fig. 11. From this figure, it is observed that the quantization performance of the proposed Duffin-CsiNet outperforms the other DL-based quantized feedback networks under different quantization bits, especially at low-resolution quantization, that illustrates the effectiveness of integrating quantization into the stage of Duffin-CsiNet training and the superiority of Duffin-CsiNet on the performance of CSI reconstruction.

On the other hand, we compare the total number of feedback bits at $\rho$ = 1/8 where the NMSE of each method is required to -22 dB, as presented in Table IV. From this table, we observe that the proposed Duffin-CsiNet can effectively save the feedback bits overhead compared with DualQnet \cite{9090892} and CH-DualNetSph-PQB \cite{9799802}.

\begin{figure}[t]
\centering
% \subfigure{
\begin{minipage}[t]{0.4\linewidth}
\centering
\includegraphics[scale=0.4]{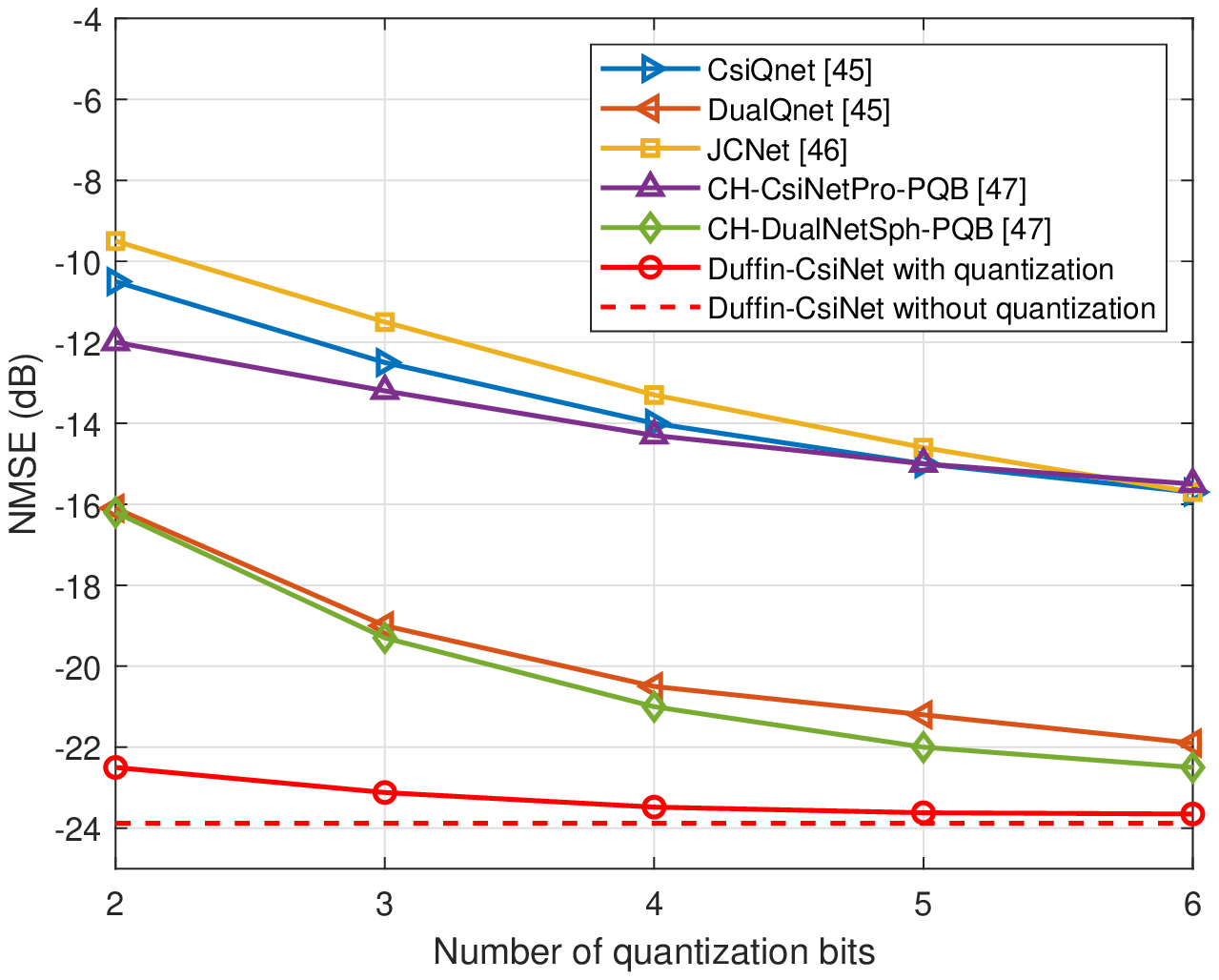}
\vspace{-1.5em}
\caption{Quantization performance comparison.}
\end{minipage}%
% }
\hspace{10pt}
% \subfigure{
\begin{minipage}[t]{0.4\linewidth}
\centering
\includegraphics[scale=0.4]{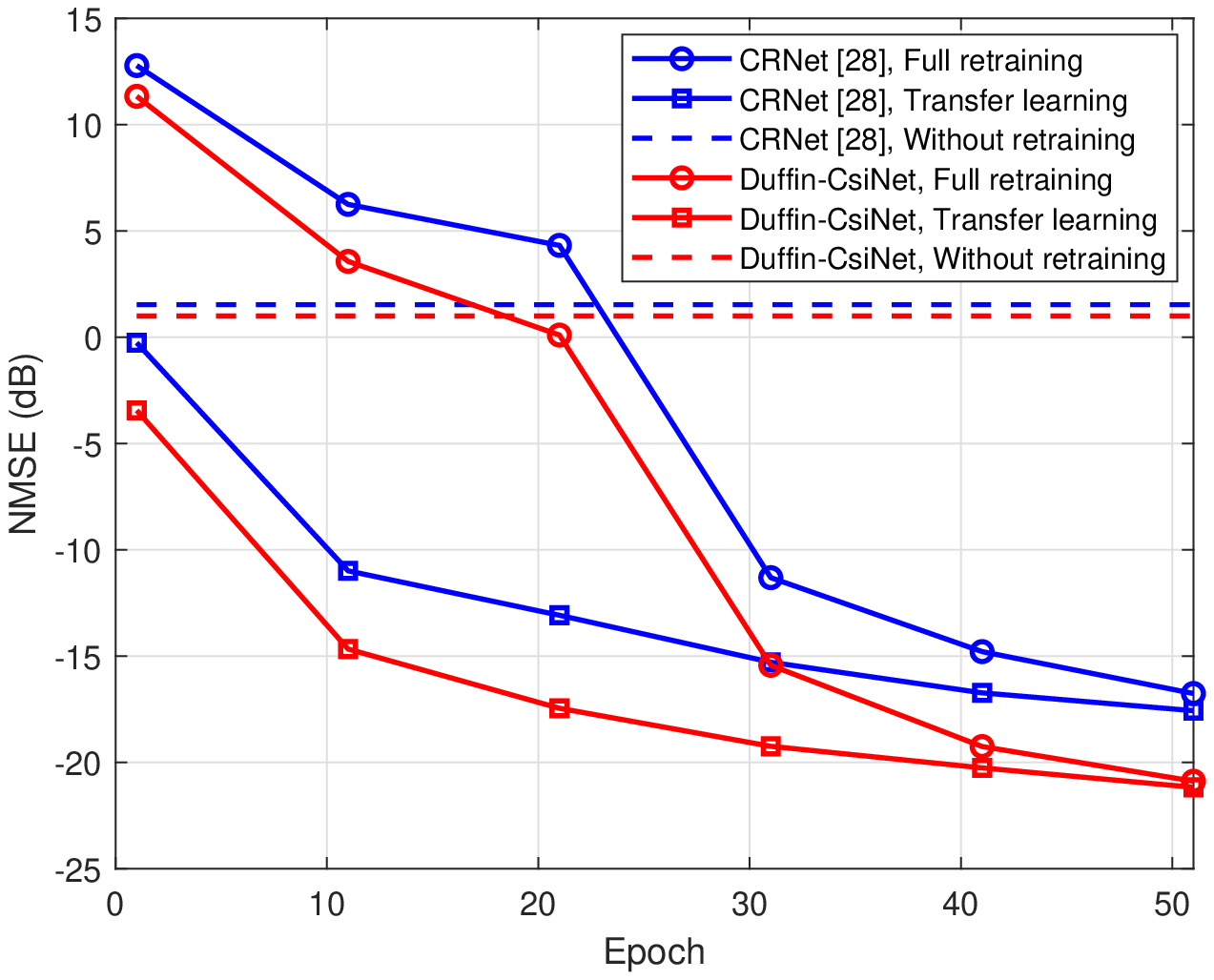}
\vspace{-1.5em}
\caption{Generalization performance comparison.}
\end{minipage}%
% }%
\end{figure}

\begin{table}[t]
  \centering
  \caption{Number comparison of feedback bits with $\rho$ = 1/8.}
  \vspace{-1.5em}
  \scalebox{0.7}{
    \begin{tabular}{cccc}
    \toprule
    Methods & \makecell{Number of \\ codeword values} & \makecell{Quantization bits \\ of per codeword} & Total bits\\
    \midrule
    DualQnet \cite{9090892} & 256   & 6     & 1,536 \\
    CH-DualNetSph-PQB \cite{9799802} & 256   & 5     & 1,280 \\
    Proposed Duffin-CsiNet & 256   & \textbf{2} & \textbf{512} \\
    \bottomrule
    \end{tabular}}%
  \label{tab:addlabel}%
\end{table}%

\subsection{Generalization}
This section analyzes the generalization of the proposed Duffin-CsiNet. Taking $\rho$ = 1/4 as an example, we test the CSI reconstruction performance of the indoor scenario using the Duffin-CsiNet trained in the outdoor scenario to discuss the generalization of Duffin-CsiNet. 
% Specifically, we first directly use the trained Duffin-CsiNet to test the NMSE of the indoor environment, as shown in the red dotted line in Fig. 12. It can be observed that the reconstruction performance of the strategy of this directly using has a large performance loss, indicating that Duffin-CsiNet cannot reconstruct CSI data with different distribution from the training set, which is also a tough problem of all current CSI compression and reconstruction NNs, e.g., \cite{9495802, 8543184, 9445070}. To solve this problem, we adopt two methods to adapt the proposed Duffin-CsiNet to the current indoor environment, i.e., the full retraining strategy and the proposed transfer learning-based strategy. 
As shown in the red curves of Fig. 12, we can observe that the proposed transfer learning-based strategy converges faster compared to the full retraining strategy. In particular, the proposed transfer learning strategy can achieve -4 dB NMSE after 1 epoch and -15 dB NMSE after 10 epochs, however the full retraining strategy requires 23 epochs and 30 epochs, respectively. This shows that the proposed transfer learning-based strategy can quickly improve the generalization of Duffin-CsiNet and enable it to perform a satisfactory reconstruction performance under a small number of network training epochs, greatly reducing the online network retraining time, which is crucial for systems with high communication latency requirement.

On the other hand, we apply the proposed transfer learning strategy to the existing CSI feedback methods, e.g., CRNet\cite{9149229}. As shown in the blue curves in Fig. 12, similar to the observation of Duffin-CsiNet, CRNet also converges faster in early epoch by using the proposed transfer learning strategy and enjoys better performance compared to full retraining. This illustrates that the proposed transfer learning strategy enjoys a powerful generalizability and can be applied to other DL-based CSI feedback methods, providing an effective solution for improving the generalization performance of DL-based CSI feedback networks.

\subsection{Complexity and Convergence}
% In practice, the network complexity is important for online using. In particular, a UE can only support a lightweight model due to its small physical size and weak signal processing capability and the BS can deploy large-scale network models thanks to the large physical size and powerful computing ability. Therefore, the complexity of the encoder should be as low as possible while without losing much performance.

In general, the NN complexity is measured by the number of network parameters, especially the trainable network parameters. Taking $\rho$ = 1/4 as an example, we compare the number of network trainable parameter between the proposed Duffin-CsiNet and other NNs in the case of indoor scenario, as shown in Table V. We observe that the encoder of the proposed Duffin-CsiNet achieve a slight parameter number reduction compared to that of other networks, which shows that the encoder of the proposed Duffin-CsiNet is suitable for the deployment at UE. In addition, due to the feature map extension of the PreNet and the cascaded structure in decoder, the decoder has a little parameter number increase compared to other methods. Fortunately, thanks to the parallel computing power of the graphics processing units (GPUs) that can be used at BS, a small increase in network trainable parameter does not greatly increase the time of online inference, i.e., without great latency increase. In particular, we present the inference time of testing set of different NNs, as shown in the third row of Table V. It can be seen that the online inference time of the proposed Duffin-CsiNet has a little increase compared to other NNs, but this little increase is marginal.

\begin{table*}[t]
\centering
\caption{Network complexity comparison, $\rho$ = 1/4.}
\vspace{-1.5em}
\scalebox{0.7}{
\begin{tabular}{c|c|c|c|c|c|c}
\hline
  & CsiNet \cite{8322184} & BCsiNet \cite{9373670} & CRNet\cite{9149229} & CLNet\cite{9497358} & ACRNet \cite{lu2021binarized} & Proposed Duffin-CsiNet \\
\hline
Encoder Param. (M) & 1.049 & 1.051 & 1.051 & 1.051 & 1.051 & \textbf{1.049} \\\hline
Decoder Param. (M) & \textbf{1.051} & 1.054 & 1.052 & 1.052 & 1.072 & 1.427 \\\hline
Inference Time (s)& \textbf{5.587} & 5.763 & 5.775 & 5.763 & 5.774 & 5.806 \\
\hline
\end{tabular}}
\end{table*}

On the other hand, we compare the convergence ability between the proposed Duffin-CsiNet and other NNs, where $\rho$ = 1/16 is applied for an example, as shown in Fig. 13. In particular, we select four CNN-based methods, including CsiNet \cite{8322184}, BCsiNet \cite{9373670}, CRNet \cite{9149229}, ACRNet \cite{lu2021binarized}, and an ANN-based method, i.e., CLNet \cite{9497358}. Specifically, we observe that the proposed Duffin-CsiNet quickly reduces the NMSE in the early epochs, e.g., from 1 to 20 epochs, and obtain better NMSE performance than other methods, i.e., below the reference line. With the epoch increasing, the proposed Duffin-CsiNet stably converges the NMSE to the minimum value, especially after 500 epochs, and finally the converged NMSE has respectively 6.85 dB gain and 3.2 dB gain compared to the CsiNet and the ACRNet. This verifies the stability and effectiveness of the proposed Duffin-CsiNet, and this stable convergence ability comes from the parallel-serial hybrid structure for feature extraction in Duffin-CsiNet and the effective strategy of learning rate change in (27).

\begin{figure}[t]
\centering
% \subfigure{
\begin{minipage}[t]{0.4\linewidth}
\centering
\includegraphics[scale=0.4]{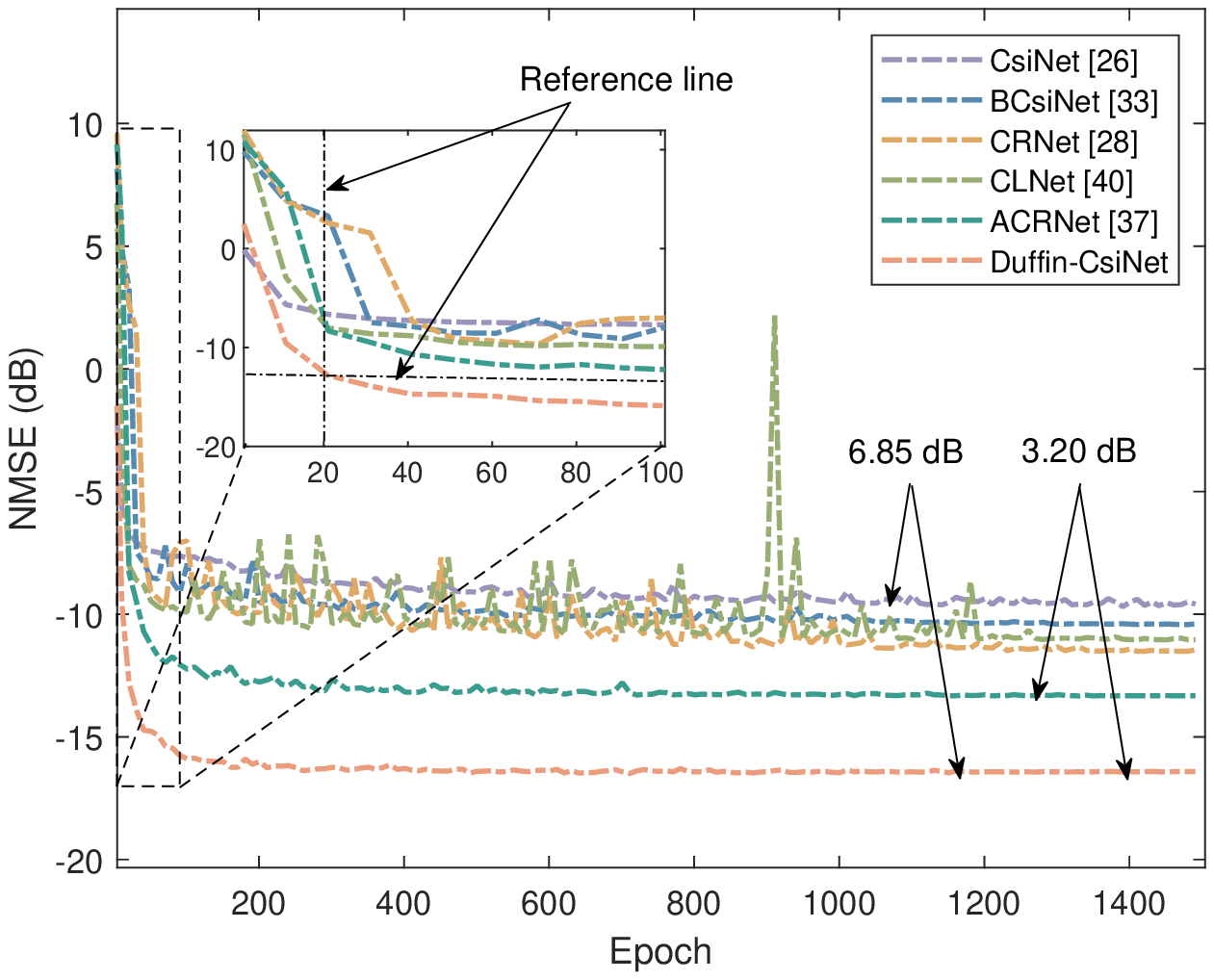}
\vspace{-1.5em}
\caption{Network convergence comparison.}
\end{minipage}%
% }
\hspace{10pt}
% \subfigure{
\begin{minipage}[t]{0.4\linewidth}
\centering
\includegraphics[scale=0.4]{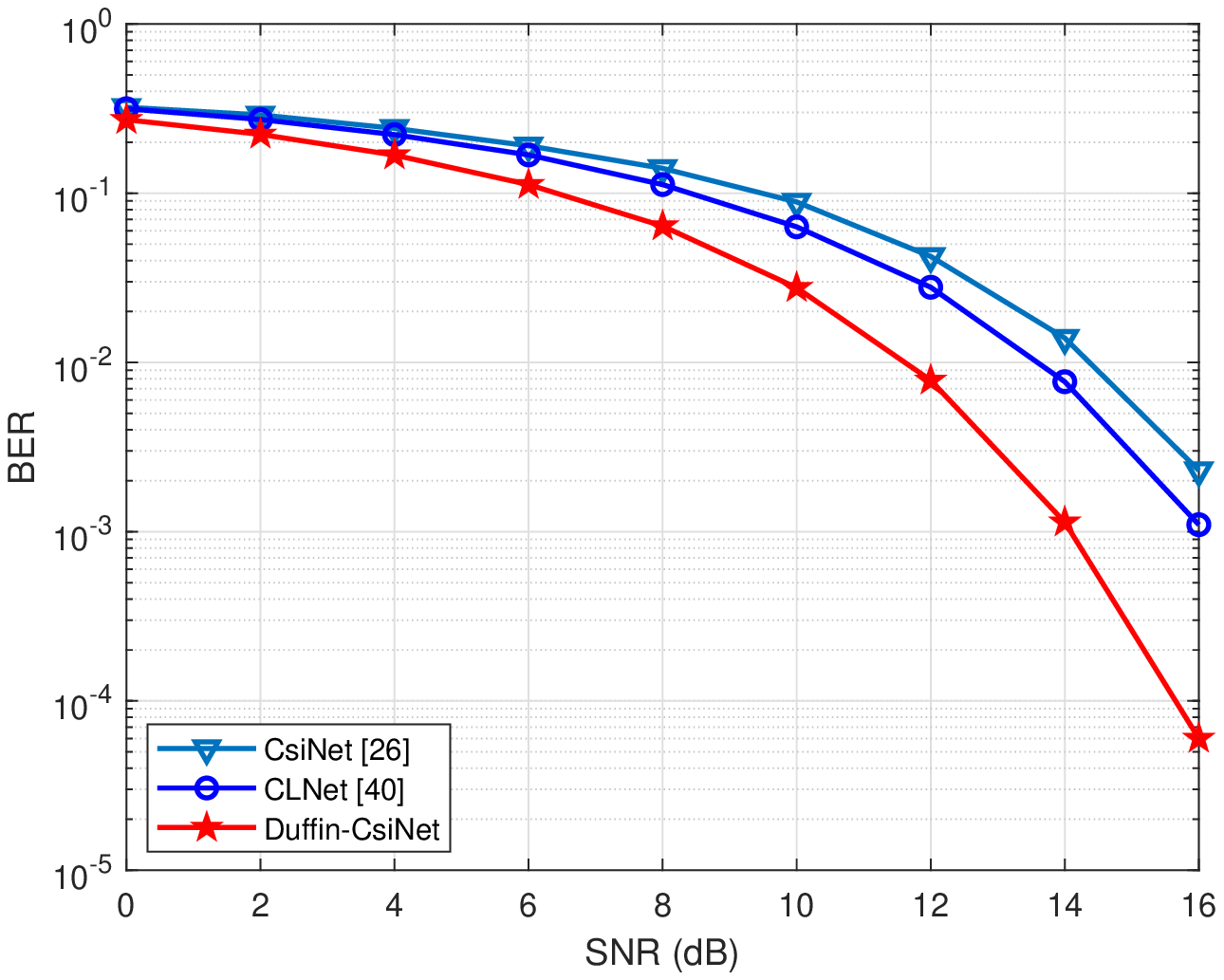}
\vspace{-1.5em}
\caption{BER performance comparison.}
\end{minipage}%
% }%
\end{figure}

% \begin{figure}[t]
%     \centering
%     \includegraphics[scale=0.35]{BER对比放论文中.eps}
%     \caption{BER performance comparison by different reconstructed CSI, $\rho$ = 1/8.}
% \end{figure}

\subsection{BER for Reconstructed CSI}

To evaluate the end-to-end performance of the system by exploiting the reconstructed CSI, we conduct the experiments on the BER of different DL-based CSI feedback methods, as presented in Fig. 14. In particular, we compare the BER performance of three CSI feedback networks, including the CsiNet \cite{8322184} focusing on the NLOS propagation-path features of the CSI image, the CLNet \cite{9497358} focusing on the dominant propagation-path features of the CSI image, and the proposed Duffin-CsiNet integrating the both features of the CSI image. Quadrature Phase Shift Keying (QPSK) modulation is adopted and the precoding vector is a maximum-ratio-transmission (MRT) beamformer designed by using the reconstructed CSI matrix. From Table II and Fig. 14, it is observed that the proposed Duffin-CsiNet achieves the lowest NMSE and outperforms the other methods in terms of BER under different signal-to-noise ratios (SNRs). This benefits from the efficient CSI feature processing module that adequately extracts the two physical features embedded in CSI image, validating the necessity for dual-propagation-feature extraction and the effectiveness of the proposed Duffin-CsiNet.

\subsection{Ablation Experiment}

In this section, we first compare the performance of the three fusion approaches in FNet, including element-wise addition, dot production, and the proposed NN-based fusion. Specifically, two scenarios including the indoor scenario and the outdoor scenario are chosen for evaluation, as presented in Table VI. As observed from the results, the proposed NN-based fusion outperforms the other two fusion approaches at the five selected compression ratios. For example, when $\rho$ = 1/4 at the indoor scenario, the proposed NN-based fusion can provide a gain of 6.23 dB and a gain of 10.85 dB compared with the element-wise addition and the dot product approaches, respectively. These results demonstrate the effectiveness of the proposed NN-based fusion.

\begin{table}[t]
  \centering
  \caption{The performance (dB) comparison of three fusion approaches.}
  \vspace{-1.5em}
  \scalebox{0.6}{
    \begin{tabular}{c|c|c|c|c|c|c}
    \toprule
    Scenario     & \diagbox{Fusion approach}{$\rho$} & 1/4 & 1/8 & 1/16 & 1/32 & 1/64 \\
    \midrule
    \multirow{3}[1]{*}{Indoor} & Element-wise addition & -28.96 & -21.34 & -15.25 & -10.63 & -7.89 \\
          & Dot product & -24.34 & -20.52 & -14.14 & -10.07 & -7.73 \\
          & Proposed NN-based fusion & \textbf{-35.19} & \textbf{-23.59} & \textbf{-17.20} & \textbf{-11.69} & \textbf{-8.05} \\
    \midrule
    \multirow{3}[1]{*}{Outdoor} & Element-wise addition & -14.88 & -10.57 & -7.09 & -4.37 & -2.96 \\
          & Dot product & -12.24 & -7.612 & -6.17 & -3.98 & -2.55 \\
          & Proposed NN-based fusion & \textbf{-16.12} & \textbf{-11.55} & \textbf{-7.84} & \textbf{-5.52} & \textbf{-3.85} \\
    \bottomrule
    \end{tabular}}%
  \label{tab:addlabel}%
\end{table}%

On the other hand, due to multipath fading in practice, the dominant propagation path usually arrives with a short delay and the NLOS propagation path arrives at with a bit longer delay \cite{8845636}. Therefore, the $N_s$ rows of $\mathbf{H}_\mathrm{d}$ is needed to be selected to ensure that all the dominant propagation-path information and most NLOS propagation-path information are retained. In general, the first $N_s$ rows with zero offsets are chosen in existing DL-based CSI feedback methods, e.g., in \cite{lu2021binarized, 9419066, yzq}. In particular, we evaluate the impact of different offsets for CSI reconstruction in the indoor scenario of COST2100 channel model \cite{6393523}, as shown in Fig. 15. The offset is selected as 0, 1, 2, 3, and 4, and the offset areas are padding by zero values after the truncated $\hat{\mathbf{H}}_\mathrm{s}$ is reconstructed. To better demonstrate the impact of the offset on the reconstructed spatial-frequency-domain CSI matrix $\mathbf{H}$, the cosine similarity $\beta$ is used for performance evaluation \cite{8322184}, which is expressed by 
\begin{equation}
    \beta =\mathbb{E}\left\{ \frac{1}{N_{\text{c}}}\sum\limits_{n=1}^{N_{\text{c}}}{\left( \dfrac{| \mathbf{\hat{h}}_n\mathbf{h}_n |}{\lVert \mathbf{\hat{h}}_n \rVert _2\lVert \mathbf{h}_n \rVert _2} \right)} \right\}. 
\end{equation}
From this figure, we observe that the cosine similarity under zero offset is the highest among the five compression ratios and the cosine similarity decreases with the increasing offset, which illustrates that the existence of offsets indeed affects the performance of CSI reconstruction and demonstrates that introducing a zero offset is the optimal for the indoor scenario of the COST2100 channel model. However, in some special scenarios, e.g., long-distance communication systems with rich scatters, the delays of all propagation paths are long and there exists an optimal non-zero offset, which may need to be derived through engineering experience or mathematical theory, providing an interesting direction for potential future research.

\begin{figure}[t] 
\centering
    \includegraphics[scale=0.6]{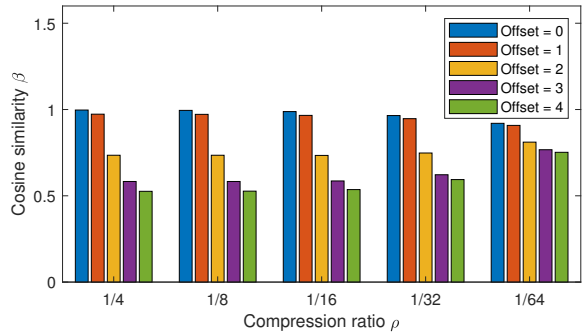}
    \vspace{-1.5em}
    \caption{Consine similarity of CSI reconstruction with different offsets in the indoor scenario of COST2100 \cite{6393523}.}
\end{figure}
% \vspace{-0.4cm}
\section{Conclusion}
This paper first proposed a dual-feature fusion enhanced network named DuffinNet for the extraction of the physical features in the CSI image. The proposed DuffinNet is a parallel-serial hybrid structure. It not only focuses on the NLOS propagation-path features through a CNN but also pays attention to the dominant propagation-path features of the CSI image through an ANN and effectively exploits their interplay through a fusion NN. 
Based on the proposed architecture of DuffinNet, this paper designed an encoder-decoder framework named Duffin-CsiNet for greatly improving the performance of MIMO CSI compression and reconstruction. 
In addition, a two-stage approach was developed for the feedback codeword quantization and a transfer learning-based method was introduced for improving the generalization of Duffin-CsiNet, which facilitates the practical deployment of the proposed Duffin-CsiNet. The simulation results showed that the proposed Duffin-CsiNet noticeably outperformed existing SOTA methods in terms of CSI reconstruction accuracy, feature extraction, encoder complexity, network convergence, and end-to-end performance of system, under various scenarios.

\bibliographystyle{IEEEtran} %这句话必须要要加
\bibliography{Areference.bib} %这句话加载bib文件

\begin{thebibliography}{10}
\providecommand{\url}[1]{#1}
\csname url@samestyle\endcsname
\providecommand{\newblock}{\relax}
\providecommand{\bibinfo}[2]{#2}
\providecommand{\BIBentrySTDinterwordspacing}{\spaceskip=0pt\relax}
\providecommand{\BIBentryALTinterwordstretchfactor}{4}
\providecommand{\BIBentryALTinterwordspacing}{\spaceskip=\fontdimen2\font plus
\BIBentryALTinterwordstretchfactor\fontdimen3\font minus
  \fontdimen4\font\relax}
\providecommand{\BIBforeignlanguage}[2]{{%
\expandafter\ifx\csname l@#1\endcsname\relax
\typeout{** WARNING: IEEEtran.bst: No hyphenation pattern has been}%
\typeout{** loaded for the language `#1'. Using the pattern for}%
\typeout{** the default language instead.}%
\else
\language=\csname l@#1\endcsname
\fi
#2}}
\providecommand{\BIBdecl}{\relax}
\BIBdecl

\bibitem{6798744}
L.~Lu, G.~Y. Li, A.~L. Swindlehurst, A.~Ashikhmin, and R.~Zhang, ``An overview
  of massive {MIMO}: Benefits and challenges,'' \emph{IEEE J. Sel. Topics
  Signal Process.}, vol.~8, no.~5, pp. 742--758, 2014.

\bibitem{7386643}
G.~N. Kamga, M.~Xia, and S.~Aïssa, ``Spectral-efficiency analysis of massive
  {MIMO} systems in centralized and distributed schemes,'' \emph{IEEE Trans.
  Commun.}, vol.~64, no.~5, pp. 1930--1941, 2016.

\bibitem{wong2017key}
V.~W. Wong, R.~Schober, D.~W.~K. Ng, and L.-C. Wang, \emph{Key Technologies for
  5G Wireless Systems}.\hskip 1em plus 0.5em minus 0.4em\relax Cambridge Univ.
  Press, 2017.

\bibitem{10024766}
W.~Xu, Z.~Yang, D.~W.~K. Ng, M.~Levorato, Y.~C. Eldar, and M.~Debbah, ``Edge
  learning for {B5G} networks with distributed signal processing: Semantic
  communication, edge computing, and wireless sensing,'' \emph{IEEE J. Sel.
  Topics Signal Process.}, vol.~17, no.~1, pp. 9--39, Jan. 2023.

\bibitem{7353214}
W.~Shen, L.~Dai, Y.~Shi, B.~Shim, and Z.~Wang, ``{Joint channel training and
  feedback for FDD massive MIMO systems},'' \emph{IEEE Trans. Veh. Technol.},
  vol.~65, no.~10, pp. 8762--8767, Oct. 2016.

\bibitem{wei2022distributed}
K.~Wei, J.~Xu, W.~Xu, N.~Wang, and D.~Chen, ``Distributed neural precoding for
  hybrid {mmWave} {MIMO} communications with limited feedback,'' \emph{IEEE
  Commun. Lett.}, vol.~26, no.~7, pp. 1568--1572, Jul. 2022.

\bibitem{9279228}
J.~Guo, C.-K. Wen, and S.~Jin, ``{Deep learning-based CSI feedback for
  beamforming in single- and multi-cell massive MIMO systems},'' \emph{IEEE J.
  Sel. Areas Commun.}, vol.~39, no.~7, pp. 1872--1884, Jul. 2021.

\bibitem{1715541}
N.~Jindal, ``{MIMO} broadcast channels with finite-rate feedback,'' \emph{IEEE
  Trans. Inform. Theory}, vol.~52, no.~11, pp. 5045--5060, 2006.

\bibitem{7470522}
J.~Park, N.~Lee, J.~G. Andrews, and R.~W. Heath, ``On the optimal feedback rate
  in interference-limited multi-antenna cellular systems,'' \emph{IEEE Trans.
  Wireless Commun.}, vol.~15, no.~8, pp. 5748--5762, 2016.

\bibitem{5529760}
L.-C. Wang and C.-J. Yeh, ``Scheduling for multiuser {MIMO} broadcast systems:
  Transmit or receive beamforming?'' \emph{IEEE Trans. Wireless Commun.},
  vol.~9, no.~9, pp. 2779--2791, Sep. 2010.

\bibitem{wang2023full}
\text{D. Wang, et al.}, ``Full-spectrum cell-free {RAN} for {6G} systems:
  system design and experimental results,'' \emph{Sci China Inf Sci}, vol.~66,
  no.~3, pp. 130\ignorespaces305:1--14, Mar. 2023.

\bibitem{8766896}
S.~Sun, N.~Akhtar, H.~Song, A.~Mian, and M.~Shah, ``{Deep affinity network for
  multiple object tracking},'' \emph{IEEE Trans. Pattern Anal. Mach. Intell.},
  vol.~43, no.~1, pp. 104--119, Jan. 2021.

\bibitem{9295376}
Y.~L. Lee, D.~Qin, L.-C. Wang, and G.~H. Sim, ``{6G} massive radio access
  networks: Key applications, requirements and challenges,'' \emph{IEEE Open J.
  Veh. Technol.}, vol.~2, pp. 54--66, Dec. 2021.

\bibitem{wei2023toward}
\text{W. Xu, et al.}, ``Toward ubiquitous and intelligent {6G} networks: from
  architecture to technology,'' \emph{Sci China Inf Sci}, vol.~66, no.~3, pp.
  130\ignorespaces300:1--2, Mar. 2023.

\bibitem{1065}
H.~Huang, Y.~Peng, J.~Yang, W.~Xia, and G.~Gui, ``{Fast beamforming design via
  deep learning},'' \emph{IEEE Trans. Veh. Technol.}, vol.~69, no.~1, pp.
  1065--1069, Jan. 2020.

\bibitem{9246287}
Q.~Hu, Y.~Cai, Q.~Shi, K.~Xu, G.~Yu, and Z.~Ding, ``{Iterative algorithm
  induced deep-unfolding neural networks: Precoding design for multiuser MIMO
  systems},'' \emph{IEEE Trans. Wireless Commun.}, vol.~20, no.~2, pp.
  1394--1410, Feb. 2021.

\bibitem{9729198}
S.~Zhang, J.~Xu, W.~Xu, N.~Wang, D.~W.~K. Ng, and X.~You, ``Data augmentation
  empowered neural precoding for multiuser {MIMO} with {MMSE} model,''
  \emph{IEEE Commun. Lett.}, vol.~26, no.~5, pp. 1037--1041, May 2022.

\bibitem{9347820}
F.~Sohrabi, K.~M. Attiah, and W.~Yu, ``{Deep learning for distributed channel
  feedback and multiuser precoding in {FDD} massive {MIMO}},'' \emph{IEEE
  Trans. Wireless Commun.}, vol.~20, no.~7, pp. 4044--4057, Jul. 2021.

\bibitem{8400482}
H.~Huang, J.~Yang, H.~Huang, Y.~Song, and G.~Gui, ``{Deep learning for
  super-resolution channel estimation and DOA estimation based massive MIMO
  system},'' \emph{IEEE Trans. Veh. Technol.}, vol.~67, no.~9, pp. 8549--8560,
  Sep. 2018.

\bibitem{8795533}
Y.~Yang, F.~Gao, G.~Y. Li, and M.~Jian, ``{Deep learning-based downlink channel
  prediction for {FDD} massive {MIMO} system},'' \emph{IEEE Commun. Lett.},
  vol.~23, no.~11, pp. 1994--1998, Nov. 2019.

\bibitem{9288911}
Q.~Hu, F.~Gao, H.~Zhang, S.~Jin, and G.~Y. Li, ``{Deep learning for channel
  estimation: Interpretation, performance, and comparison},'' \emph{IEEE Trans.
  Wireless Commun.}, vol.~20, no.~4, pp. 2398--2412, Apr. 2021.

\bibitem{9075976}
J.~Liao, J.~Zhao, F.~Gao, and G.~Y. Li, ``{A model-driven deep learning method
  for massive MIMO detection},'' \emph{IEEE Commun. Lett.}, vol.~24, no.~8, pp.
  1724--1728, Aug. 2020.

\bibitem{9018199}
H.~He, C.-K. Wen, S.~Jin, and G.~Y. Li, ``{Model-driven deep learning for MIMO
  detection},'' \emph{IEEE Trans. Signal Process.}, vol.~68, pp. 1702--1715,
  Feb. 2020.

\bibitem{9517121}
R.~Xie, W.~Xu, Y.~Chen, J.~Yu, A.~Hu, D.~W.~K. Ng, and A.~Lee~Swindlehurst,
  ``{A generalizable model-and-data driven approach for open-set RFF
  authentication},'' \emph{IEEE Trans. Inf. Forensics Security}, vol.~16, pp.
  4435--4450, Aug. 2021.

\bibitem{7166317}
B.~Lee, J.~Choi, J.-Y. Seol, D.~J. Love, and B.~Shim, ``{Antenna grouping based
  feedback compression for FDD-based massive MIMO systems},'' \emph{IEEE Trans.
  Commun.}, vol.~63, no.~9, pp. 3261--3274, Sep. 2015.

\bibitem{8322184}
C.-K. Wen, W.-T. Shih, and S.~Jin, ``{Deep learning for massive {MIMO} {CSI}
  feedback},'' \emph{IEEE Wireless Commun. Lett.}, vol.~7, no.~5, pp. 748--751,
  Oct. 2018.

\bibitem{9178295}
X.~Yu, X.~Li, H.~Wu, and Y.~Bai, ``{{DS-NLCsiNet}: Exploiting non-local neural
  networks for massive {MIMO CSI} feedback},'' \emph{IEEE Commun. Lett.},
  vol.~24, no.~12, pp. 2790--2794, Dec. 2020.

\bibitem{9149229}
Z.~Lu, J.~Wang, and J.~Song, ``{Multi-resolution {CSI} feedback with deep
  learning in massive {MIMO} system},'' in \emph{Proc. IEEE Int. Conf. Commun.
  (ICC)}, Jun. 2020, pp. 1--6.

\bibitem{He_2016_CVPR}
K.~He, X.~Zhang, S.~Ren, and J.~Sun, ``{Deep residual learning for image
  recognition},'' in \emph{Proc. Conf. Comput. Vis. Pattern Recognit (CVPR)},
  Jun. 2016, pp. 770--778.

\bibitem{2004}
I.~Daubechies, M.~Defrise, and C.~De~Mol, ``{An iterative thresholding
  algorithm for linear inverse problems with a sparsity constraint},''
  \emph{Commun. Pure Appl. Math.}, vol.~57, no.~11, pp. 1413--1457, Nov. 2004.

\bibitem{7457256}
C.~A. Metzler, A.~Maleki, and R.~G. Baraniuk, ``{From denoising to compressed
  sensing},'' \emph{IEEE Trans. Inf. Theory}, vol.~62, no.~9, pp. 5117--5144,
  Sep. 2016.

\bibitem{9495802}
Z.~Hu, J.~Guo, G.~Liu, H.~Zheng, and J.~Xue, ``{MRFNet: A deep learning-based
  CSI feedback approach of massive MIMO systems},'' \emph{IEEE Commun. Lett.},
  vol.~25, no.~10, pp. 3310--3314, Jul. 2021.

\bibitem{9373670}
Z.~Lu, J.~Wang, and J.~Song, ``{Binary neural network aided CSI feedback in
  massive MIMO system},'' \emph{IEEE Wireless Commun. Lett.}, vol.~10, no.~6,
  pp. 1305--1308, Jun. 2021.

\bibitem{9552908}
B.~Cao, Y.~Yang, P.~Ran, D.~He, and G.~He, ``{ACCsiNet}: Asymmetric
  convolution-based autoencoder framework for massive {MIMO} {CSI} feedback,''
  \emph{IEEE Commun. Lett.}, vol.~25, no.~12, pp. 3873--3877, 2021.

\bibitem{8543184}
C.~Lu, W.~Xu, H.~Shen, J.~Zhu, and K.~Wang, ``{MIMO channel information
  feedback using deep recurrent network},'' \emph{IEEE Commun. Lett.}, vol.~23,
  no.~1, pp. 188--191, Jan. 2019.

\bibitem{9445070}
X.~Song, J.~Wang, J.~Wang, G.~Gui, T.~Ohtsuki, H.~Gacanin, and H.~Sari,
  ``{{SALDR}: Joint self-attention learning and dense refine for massive {MIMO
  CSI} feedback with multiple compression ratio},'' \emph{IEEE Wireless Commun.
  Lett.}, vol.~10, no.~9, pp. 1899--1903, Jun. 2021.

\bibitem{lu2021binarized}
Z.~Lu, X.~Zhang, H.~He, J.~Wang, and J.~Song, ``Binarized aggregated network
  with quantization: Flexible deep learning deployment for {CSI} feedback in
  massive {MIMO} systems,'' \emph{IEEE Trans. Wireless Commun.}, vol.~21,
  no.~7, pp. 5514--5525, Jul. 2022.

\bibitem{9419066}
Z.~Cao, W.-T. Shih, J.~Guo, C.-K. Wen, and S.~Jin, ``{{Lightweight
  convolutional neural networks for {CSI} feedback in massive {MIMO}}},''
  \emph{IEEE Commun. Lett.}, vol.~25, no.~8, pp. 2624--2628, Aug. 2021.

\bibitem{yzq}
Z.~Yin, W.~Xu, R.~Xie, S.~Zhang, D.~W.~K. Ng, and X.~You, ``Deep {CSI}
  compression for massive {MIMO}: A self-information model-driven neural
  network,'' \emph{IEEE Trans. Wireless Commun.}, vol.~21, no.~10, pp.
  8872--8886, Oct. 2022.

\bibitem{9497358}
S.~Ji and M.~Li, ``{CLNet}: Complex input lightweight neural network designed
  for massive {MIMO} {CSI} feedback,'' \emph{IEEE Wireless Commun. Lett.},
  vol.~10, no.~10, pp. 2318--2322, Oct. 2021.

\bibitem{8972904}
J.~Guo, C.-K. Wen, S.~Jin, and G.~Y. Li, ``{Convolutional neural network-based
  multiple-rate compressive sensing for massive {MIMO CSI} feedback: Design,
  simulation, and analysis},'' \emph{IEEE Trans. Wireless Commun.}, vol.~19,
  no.~4, pp. 2827--2840, Apr. 2020.

\bibitem{9171358}
Y.~Sun, W.~Xu, L.~Fan, G.~Y. Li, and G.~K. Karagiannidis, ``{AnciNet: An
  efficient deep learning approach for feedback compression of estimated {CSI}
  in massive {MIMO} systems},'' \emph{IEEE Wireless Commun. Lett.}, vol.~9,
  no.~12, pp. 2192--2196, Dec. 2020.

\bibitem{9439959}
Y.~Sun, W.~Xu, L.~Liang, N.~Wang, G.~Y. Li, and X.~You, ``{A lightweight deep
  network for efficient {CSI} feedback in massive {MIMO} systems},'' \emph{IEEE
  Wireless Commun. Lett.}, vol.~10, no.~8, pp. 1840--1844, Aug. 2021.

\bibitem{9296555}
M.~B. Mashhadi, Q.~Yang, and D.~Gündüz, ``Distributed deep convolutional
  compression for massive {MIMO} {CSI} feedback,'' \emph{IEEE Trans. Wireless
  Commun.}, vol.~20, no.~4, pp. 2621--2633, 2021.

\bibitem{9090892}
Z.~Liu, L.~Zhang, and Z.~Ding, ``An efficient deep learning framework for low
  rate massive {MIMO} {CSI} reporting,'' \emph{IEEE Trans. Commun.}, vol.~68,
  no.~8, pp. 4761--4772, 2020.

\bibitem{8845636}
C.~Lu, W.~Xu, S.~Jin, and K.~Wang, ``{Bit-level optimized neural network for
  multi-antenna channel quantization},'' \emph{IEEE Wireless Commun. Lett.},
  vol.~9, no.~1, pp. 87--90, Jan. 2020.

\bibitem{9799802}
X.~Liang, H.~Chang, H.~Li, X.~Gu, and L.~Zhang, ``Changeable rate and novel
  quantization for {CSI} feedback based on deep learning,'' \emph{IEEE Trans.
  Wireless Commun.}, vol.~21, no.~12, pp.
  10\ignorespaces100--10\ignorespaces114, 2022.

\bibitem{1033686}
A.~M. Sayeed, ``{Deconstructing multiantenna fading channels},'' \emph{IEEE
  Trans. Signal Process.}, vol.~50, no.~10, pp. 2563--2579, Oct. 2002.

\bibitem{6393523}
L.~Liu, C.~Oestges, J.~Poutanen, K.~Haneda, P.~Vainikainen, F.~Quitin,
  F.~Tufvesson, and P.~D. Doncker, ``{The {COST} 2100 {MIMO} channel model},''
  \emph{IEEE Wireless Commun.}, vol.~19, no.~6, pp. 92--99, Dec. 2012.

\bibitem{7368949}
U.~Ugurlu, R.~Wichman, C.~B. Ribeiro, and C.~Wijting, ``A multipath
  extraction-based {CSI} acquisition method for {FDD} cellular networks with
  massive antenna arrays,'' \emph{IEEE Trans. Wireless Commun.}, vol.~15,
  no.~4, pp. 2940--2953, Apr. 2016.

\bibitem{woo2018cbam}
S.~Woo, J.~Park, J.-Y. Lee, and I.~S. Kweon, ``{CBAM: Convolutional block
  attention module},'' in \emph{Proc. Eur. Conf. Comput. Vis. (ECCV)}, Sep.
  2018, pp. 3--19.

\bibitem{9156697}
Q.~Wang, B.~Wu, P.~Zhu, P.~Li, W.~Zuo, and Q.~Hu, ``{ECA-Net: Efficient channel
  attention for deep convolutional neural networks},'' in \emph{Proc. Conf.
  Comput. Vis. Pattern Recognit (CVPR)}, Jun. 2020, pp.
  11\ignorespaces531--11\ignorespaces539.

\bibitem{8935405}
W.~Xia, G.~Zheng, Y.~Zhu, J.~Zhang, J.~Wang, and A.~P. Petropulu, ``{A deep
  learning framework for optimization of MISO downlink beamforming},''
  \emph{IEEE Trans. Commun.}, vol.~68, no.~3, pp. 1866--1880, Apr. 2020.

\bibitem{8461983}
N.~Farsad, M.~Rao, and A.~Goldsmith, ``Deep learning for joint source-channel
  coding of text,'' in \emph{Proc. IEEE Int. Conf. Acoust. Speech, Signal
  Process. (ICASSP)}, 2018, pp. 2326--2330.

\bibitem{9057584}
X.~Xiao, B.~Vasić, R.~Tandon, and S.~Lin, ``Designing finite alphabet
  iterative decoders of {LDPC} codes via recurrent quantized neural networks,''
  \emph{IEEE Trans. Commun.}, vol.~68, no.~7, pp. 3963--3974, 2020.

\end{thebibliography}

\end{document}